\documentclass{article}
\usepackage[utf8]{inputenc}

\usepackage{hyperref}
\usepackage{amsmath,amsfonts,amssymb}
\usepackage{amsthm}
\usepackage{color}
\usepackage{graphicx}
\usepackage{tikz-cd}
\usepackage{tikz-3dplot}
\usetikzlibrary {decorations.pathmorphing,shadows,fadings}
\usepackage{pgfplots}
\usepackage{cite}

\def\ds{\displaystyle}
\def\bea{\begin{array}{c}}
\def\ea{\end{array}}
\def\be{\begin{equation}\bea\ds}
\def\ee{\ea\end{equation}}
\def\bee{\begin{equation}\begin{array}{rcl}\ds}
\def\eee{\end{array}\end{equation}}

\def\tr{{\rm Tr}\,}
\def\nn{\nonumber}
\def\Lc{{\mathcal{L}}}
\def\Hc{{\mathcal{H}}}

\def\Mc{{\mathcal{M}}}

\def\tr{{\rm tr}\,}
\def\Tr{{\rm Tr}\,}

\def\nn{\nonumber}

\title{Connectomes as Holographic States}
\author{Dmitry Melnikov}
\date{}

\begin{document}

\maketitle

\begin{center}
\textit{\small International Institute of Physics, Federal University of 
Rio Grande do Norte, \\ Campus Universit\'ario, Lagoa Nova, Natal-RN  
59078-970, Brazil}

\vspace{2cm}

\end{center}

\vspace{-2cm}

\begin{abstract}
     We use the topological quantum field theory description of states in Chern-Simons theory to discuss the relation between spacetime connectivity and entanglement, exploring the paradigm entanglement=topology. We define a special class of states in Chern-Simons with properties similar to those of holographic states. While the holographic states are dual to classical geometries, these connectome states represent classical topologies, which satisfy a discrete analog of the Ryu-Takayanagi formula and characteristic inequalities for the entanglement entropy. Generic states are linear combinations of connectomes, and the theory also has nonperturbative states which are global spacetime defects formed by a large number of quantum fluctuations. Topological presentation of quantum states and emergence of topology from entanglement may be useful for building a generalization to geomentry, that is quantum gravity.    Thinking of further quantum gravity comparisons we discuss replica wormholes and conclude that similar objects exist beyond gravitational theories. The topological theory perspective suggests that the sum over all wormholes is always factorizable, even though the individual ones might not be.
\end{abstract}

\bigskip

\section{Introduction}

Chern-Simons theory is an example of topological quantum field theories (TQFT) which map topological spaces into instances of quantum mechanics~\cite{Witten:1988ze,Atiyah:1989vu}. In the case of a compact gauge group it is also an example of fully solvable quantum field theory, in which any given correlation function can be explicitly computed~\cite{Witten:1988hf}. The price paid for the solvability is the fact that the theory does not have dynamical degrees of freedom, which makes it too trivial for addressing important dynamical phenomena. Nevertheless, Chern-Simons theory is not completely trivial, and the purpose of this paper is to revisit some basic facts and techniques in order to offer a complementary perspective on states in quantum gravity and the emergence of spacetime.

The specific aspects that will be considered in this paper are related to holography~\cite{Maldacena:1997re,Gubser:1998bc,Witten:1998qj}, and in particular, to the properties of quantum states described by semiclassical and classical gravity. Loosely speaking holographic duality conjectures an equivalence between quantum gravity or strings in an anti de Sitter kind of spacetime and a conformal field theory (CFT) on the boundary of that space. In the regime, in which the spacetime curvature is small, gravity, which becomes classical, probes the strongly coupled regime of the dual CFT. Among the questions that are commonly asked in the perspective of this regime are how general the quantum states described by classical gravity are, how quantum corrections to classical gravity state are computed and what happens with the spacetime in the quantum regime, e.g.~\cite{VanRaamsdonk:2010pw,Swingle:2012wq,Bao:2015bfa}. Here we will touch upon similar questions about states and Chern-Simons theory with a compact group~\cite{Witten:1988hf} and observe properties analogous to that of holographic states. In particular we will propose a set of states (topologies) in Chern-Simons theory that are analogs of classical gravity states or classical geometries. 

The main motivation to compare Chern-Simons states and gravity states is the fact that classical gravity in three dimensions has a formulation as a Chern-Simons theory with noncompact group $SL(2,R)\times SL(2,R)$~\cite{Achucarro:1986uwr,Witten:1988hc}. The status of this correspondence remains unclear at the quantum level, since no consensus is reached as of yet about the existence of a well defined quantum versions of either of the two theories~\cite{Maloney:2007ud,Maloney:2020nni}. Compact group Chern-Simons looks interesting in this respect since it may share certain properties with the noncompact version, and besides, is fully computable. Moreover, the compact group Chern-Simons is also holographic, since it has an equivalent description in terms of states of two-dimensional Wess-Zumino-Witten (WZW) theory on the boundary~\cite{Witten:1988hf}.

The class of Chern-Simons states that will be the main focus of the paper are defined in terms of open Wilson lines connecting two-dimensional boundaries of three-dimensional spaces. In TQFT open spaces correspond to states in Hilbert spaces determined by the topology of space's boundary. In particular, disconnected boundaries correspond to tensor products of the respective Hilbert spaces~\cite{Atiyah:1989vu}. Given the distribution of the endpoints of the Wilson lines on the boundaries we will be concerned with the situation which realizes the simplest topology of the connection of the endpoints in the bulk~\cite{Melnikov:2022qyt}. Equivalently, we would like to think of the states as defined by the adjacency matrix of a graph, whose nodes correspond to the disconnected components of the boundary, and edges -- to the Wilson lines connecting the boundaries. As in the previous work we will refer to these states as \emph{connectomes} owing the terminology to similar graphs in neuroscience.

The definition of connectomes above is ambiguous, since it is not in general obvious which topology is the ``simplest''. For a bipartite system (with two boundary components) the simplest topology can be expressed by planar graphs. However, the planar graph definition cannot be extended to an arbitrary number of parties. It turns out that there is an unambiguous way of defining the simplest topologies in terms of the classical limit of the Chern-Simons theory. In this limit the simplest topologies appear as equivalence classes of local braiding of the Wilson lines. Consequently, we will call connectomes those quantum states that survive the classical limit. We can thus also call them \emph{classical topologies}, by analogy with classical geometries.

We will show that classical topologies share similar properties with classical geometries. One such property is the Ryu-Takayanagi (RT) formula, which calculates the von Neumann entropy of a subregion on the boundary in terms of a minimal area surface in the bulk. We will show that connectome states satisfy a discrete version of the RT formula, similar to the one appearing in the tensor network construction of holography~\cite{Pastawski:2015qua} or in the bit-thread models~\cite{Freedman:2016zud,Headrick:2022nbe}.  Away from the classical regime, the RT formula receives corrections, whose form can be obtained from the full solution.

Besides the entropy formula, connectome states apparently satisfy the same inequalities for the entanglement entropy as the classical geometries. Such inequalities were extensively studied in the literature: see~\cite{Headrick:2007km,Hayden:2011ag,Bao:2015bfa} for some original proposals and~\cite{Czech:2023xed} for the current state of the art and references. Similar inequalities for the connectome states were discussed in~\cite{Melnikov:2023nzn}, where a number of examples were checked, all of them satisfying the identities (saturating them in some cases). It was then conjectured that connectomes satisfy the same inequalities as the holographic states, which implies that the former constitute a subclass of the latter.

We will then see that connectomes are ``perturbative'' classical topologies, since there are also nonconnectome topologies that survive the classical limit. These ``nonperturbative'' topologies are characterized by three-dimensional defects introduced in the bulk. Employing the surgery operation we will see that classical topologies with defects are obtained as a sum over a large number of perturbative topologies.

The findings of this paper are consistent with a picture of emergence of a discrete bulk spacetime from entanglement.\footnote{In fact, Chern-Simons theory provides a realization of Penrose's spin network~\cite{Penrose:1971ang}, which is a discrete model of emergent spacetime.} In order to have the space connected, one needs to have Wilson lines, which ensure that quantum states are entangled. Nonconnectome topologies describe fluctuations of the spacetime, which can, among other things, forge links between the disconnected pieces. Such \emph{quantum topologies} are themselves linear combinations of the classical ones.

As one development of the last property we will also apply the Chern-Simons perspective to the problem of replica wormholes in gravity theories. It was previously observed that inclusion of certain topologies in the calculation of semiclassical path integrals of gravity by the replica method helps solve different paradoxes in black hole physics~\cite{Penington:2019kki,Almheiri:2019qdq}. A notorious case is the Hawking information paradox~\cite{Hawking:1976ra}, in which classical saddles corresponding to a nontrivial topology connecting different replicas are shown to ensure the restoration of information at late times of the black hole evolution~\cite{Penington:2019npb,Almheiri:2019psf,Almheiri:2019hni,Almheiri:2019qdq}.

Since the replica wormhole story involves summing over topologies one might expect that TQFT should provide basic examples of the corresponding mechanism. We will arrive to the conclusion that replica wormholes are not specific to gravitational theories. Moreover, the TQFT perspective suggests that relevant wormholes are always factorizable in the full quantum picture, although their contribution may appear as a sum of non-factorizable classical topologies. Finally, the change of dominance of different replicas in the path integral is a matter of choice of an appropriate evolution Hamiltonian, external to the TQFT. So, as far as the black hole information paradox is concerned, replica wormholes are only consequences of the choice of the Hamiltonian rather than themselves solutions to the paradox.

The paper is organized as follows. In section~\ref{sec:chernsimons} we give a minimal review of the necessary details about states in Chern-Simons theory. In section~\ref{sec:basicCS} we review the basic necessary information to work with quantum mechanics based on Chern-Simons theory. Section~\ref{sec:topoqubits} sets the stage for quantum information applications,  introduces the class of connectome states and discusses their properties. In section~\ref{sec:entropy} we revisit the entanglement entropy calculation: for connectome states in section~\ref{sec:minarrea}, where we demonstrate the validity of the RT formula; and for generic states in~\ref{sec:corrections}, where we give some examples of corrections to the RT formula. In section~\ref{sec:classlimit} we discuss the classical limit of the Chern-Simons states and show that all perturbative states reduce to connectomes in the classical limit. In section~\ref{sec:wormholes} we discuss non-perturbative states, i.e. states with global defects in the three-dimensional topology. We also discuss the RT formula in the presence of such defects. In section~\ref{sec:replicawormholes} we discuss replica wormholes in the context of Chern-Simons theory. In section~\ref{sec:conclusions} we give another summary of the observations and results obtained in this paper.

\section{Connectome states in Chern-Simons theory}
\label{sec:chernsimons}

\subsection{Basics of Chern-Simons TQFT}
\label{sec:basicCS}

Let us briefly introduce a specific quantum mechanical system, which realizes a TQFT. We will focus on the example of $SU(2)_k$ Chern-Simons, but the discussion can be generalized to other examples of TQFT. More details about the specific construction can be found in~\cite{Melnikov:2018zfn,Melnikov:2022qyt}. For formal introduction of TQFT see~\cite{Witten:1988ze,Atiyah:1989vu}.

Quantum states of 3-dimensional Chern-Simons theory are represented by 3-manifolds with 2-dimensional boundaries. An example of such states are the following two vectors:
\be
\label{CSstates}
|\hat{0}\rangle \ = \  \begin{array}{c}
     \includegraphics[scale=0.15]{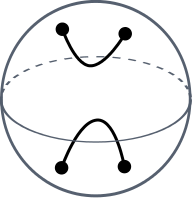} 
\end{array} \ \equiv \begin{array}{c}\scalebox{0.6}{
\begin{tikzpicture}[very thick]
\fill[black] (0,0.0) circle (0.05cm);
\fill[black] (0,0.3) circle (0.05cm);
\fill[black] (0,0.6) circle (0.05cm);
\fill[black] (0,0.9) circle (0.05cm);
\draw (0,0) -- (0.4,0) arc (-90:90:0.15cm) -- (0,0.3);
\draw (0,0.6) -- (0.4,0.6) arc (-90:90:0.15cm) -- (0,0.9);
\end{tikzpicture}} 
\end{array}\,,\qquad |\hat 1\rangle \ = \ \begin{array}{c}
     \includegraphics[scale=0.15]{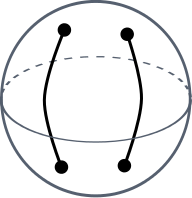} 
\end{array}  \equiv \ \begin{array}{c}\scalebox{0.6}{
\begin{tikzpicture}[very thick]
\fill[black] (0,0.0) circle (0.05cm);
\fill[black] (0,0.3) circle (0.05cm);
\fill[black] (0,0.6) circle (0.05cm);
\fill[black] (0,0.9) circle (0.05cm);
\draw (0,0) -- (0.1,0) arc (-90:90:0.45cm) -- (0,0.9);
\draw (0,0.6) -- (0.1,0.6) arc (90:-90:0.15cm) -- (0,0.3);
\end{tikzpicture}} 
\end{array}\,.
\ee
In this example the 2-manifold is a 2-sphere and the 3-manifold is a 3-ball filling the sphere. The spheres have punctures, and the punctures are extended into three-dimensional bulk as Wilson lines. In what follows we will consider Wilson lines in the fundamental representation of $SU(2)$. The above example also illustrates that connecting punctures in inequivalent ways in general produces linearly independent states in the Hilbert space.

That manifolds of~(\ref{CSstates}) represent states follows from the axioms of TQFT~\cite{Witten:1988ze,Atiyah:1989vu}. The same can be understood in the path integral definition,
\be
 |\Psi(\Sigma,\gamma_j,R_j)\rangle \ = \ \int {\mathcal D}A\Big|_{A(\Sigma)=A_\Sigma} {\rm e}^{iS_{\rm CS}[\Mc,\Sigma,\gamma_j,R_j]}\,.
\ee
Here the path integral is calculated over a 3-manifold $\Mc$ with boundary $\Sigma$, with prescribed boundary conditions $A_\Sigma$ for the gauge fields on the boundary. The Chern-Simons theory action may contain insertions of open and closed Wilson lines $\gamma_j$ in representation $R_j$.

Axioms and the path integral realization readily explain how the inner product in the Hilbert space is computed. One must glue two 3-manifolds over a common boundary, so that the result is a (path integral over) closed 3-manifold, i.e. a $\mathbb{C}$-number. The result is non-trivial only if the Wilson line ending on the boundary of two 3-manifold match, so that all Wilson lines of the closed 3-manifold are closed. This is generalized to the definitions of linear operators, tensors, traces and contractions in a straightforward way.

Matching the Wilson lines in the internal product means that both factors belong to the same Hilbert space (or rather to a Hilbert space and its dual). This Hilbert space is completely determined by the boundary $\Sigma$ (including the punctures). For $n$ punctures this Hilbert space is isomorphic to the space of $n$-point conformal blocks of $SU(2)_k$ WZW theory. In particular the dimension of this Hilbert space is given by the number of singlets in a tensor product of the representation of the $n$ punctures. A simple analogy is the requirement that the total $SU(2)$ spin of the punctures should be zero. For $n$ spin-half punctures this is given by the Catalan number
\be
\label{dimH}
\dim {\Hc_n} \ = \ C_{n/2} \ = \ \frac{(n)!}{((n/2)+1)!(n/2)!}\,, \qquad k>n/2\,.
\ee
There is an important subtlety here. The counting must be applied not to regular $SU(2)$ irreps but to integrable representations of $SU(2)_k$ WZW theory. In the latter available spin $j$ representations are limited by $j\leq k/2$. Higher spin representation simply do not exist. This implies that for general $k$ $\dim {\Hc_n}  \leq  C_{n/2}$. One of the consequences is that states given by different topology might be linearly dependent. However, if one works with sufficiently large $k$, as equation~(\ref{dimH}) suggests, the above counting is correct. In the paper we will in fact be interested in the limit $k\to\infty$.

In examples~(\ref{CSstates}) we used a single $S^2$ as the boundary $\Sigma$. However, $\Sigma$ can be any two-dimensional surface, as in the case of knot complement states studied in~\cite{Salton:2016qpp,Balasubramanian:2016sro,Balasubramanian:2018por}, or any disjoint union of two-dimensional surfaces. Here we will focus on states given by $\Sigma$ that is a collection of non-intersecting $S^2$ boundaries. To simplify the diagrams we will not try to draw 3-manifolds, nor their boundary spheres. The location of the boundaries will instead be indicated by grouping the endpoints of the Wilson line, as in equation~(\ref{CSstates}).

Finally, we will need a technology to compute the overlap of states, or more generally, correlation functions. As is famously known~\cite{Witten:1988hf} correlation functions of Wilson loop operators in $SU(2)$ Chern-Simons theory are given by the Jones polynomials~\cite{Jones:1985dw} of the knots (links) following the Wilson loop paths, for which various methods of computation exist. We will review the simplest prescription to compute them using the variation of Conway skein relations due to Kauffman~\cite{Kauffman:1987sta}.

To compute the Jones polynomial of a knot or a link in $S^3$, that is to compute an overlap of two states glued together to form a 3-sphere (which is the result of joining a 3-ball and another 3-ball turned inside out) we will use the following three rules:
\begin{itemize}
    \item The polynomial of a single unknotted ring is simply a number $d$,
    \be
    J(\begin{tikzpicture}[thick]
\draw[black] (0,0.0) circle (0.15cm);
\end{tikzpicture})\ =\ d\,;
    \ee
    \item The polynomial of a disjoint union of a knot (link) $\Lc$ and an unknotted circle (that is  circle topologically disconnected from $\Lc$) is given by
    \be
    J(\begin{tikzpicture}[thick]
    \draw[black] (0,0.0) circle (0.15cm);
\end{tikzpicture} 
\cup \Lc) \ = \ d\cdot J(\Lc)
    \ee
that is the polynomial of unlinked components factorizes.
    \item Finally, the knot (link) may be nontrivial, but the polynomial still can be computed by recursive application of the skein relation~\cite{Kauffman:1987sta}. In particular, any crossing of the knot diagram can be ``resolved'' by replacing the original overlap (path integral) by a linear combination of two overlaps, in which that specific crossing is changed by two possible ways of connecting the lines without crossing. Specifically, the following is true for the knot diagrams:
\be
\label{skein}
\begin{tikzpicture}[baseline=0]
\draw[thick] (0.5,-0.1) -- (0.3,-0.1) -- (0.15,0);
\draw[thick] (-0.15,0.2) -- (-0.3,0.3) -- (-0.5,0.3);
\draw[thick] (-0.5,-0.1) -- (-0.3,-0.1) -- (0.3,0.3) -- (0.5,0.3);
\end{tikzpicture}  \quad = \ A\quad \begin{tikzpicture}[baseline=0]
\draw[thick] (-0.5,-0.1) -- (-0.3,-0.1) arc (-90:90:0.2) -- (-0.5,0.3);
\draw[thick] (0.5,-0.1) -- (0.3,-0.1) arc (-90:-270:0.2) -- (0.5,0.3);
\end{tikzpicture} \quad +\ A^{-1}\quad \begin{tikzpicture}[baseline=0]
\draw[thick] (0.5,-0.1) -- (-0.5,-0.1);
\draw[thick] (0.5,0.3) -- (-0.5,0.3);
\end{tikzpicture}
\ee
A simple self-consistency check shows that parameters $d$ and $A$ must be related:
\be
d \ = \ -A^2 - A^{-2}\,.
\ee
Recursive resolution of all the crossings reduce the knot diagram to a linear combination of diagrams containing only unlinked circles and the remaining two rules apply.
    \end{itemize}

The above prescription produces a noncanonical normalization of Jones polynomials. In particular, the standard normalization of the Jones polynomial of the unknotted circle is unity. Polynomials computed above will always have an extra factor of $d$, as compared to the standard normalization. The prescription also distinguishes nonplanar isotopies of the knots, which produce powers of $-A^3$. This ambiguity results in an ambiguous phase of quantum states. In Chern-Simons theory it is known as the problem of framing~\cite{Witten:1988hf}. To finalize the connection with Chern-Simons we define $A$ in terms of the Chern-Simons data:
\be
A \ = \ \exp\left(\frac{\pi i}{2(k+2)}\right).
\ee

Now we are all set to discuss specific properties of quantum states in Chern-Simons theory.

\subsection{Topological bits and connectomes}
\label{sec:topoqubits}

What is a qubit in the present Chern-Simons formulation? Note that for $n=4$ equation~(\ref{dimH}) gives $\dim\Hc \ = \ 2$, so states in formula~(\ref{CSstates}) are states in a two-dimensional Hilbert space. Moreover they are linearly independent (with a caveat discussed below). One can use the rules of computing overlaps from the previous section to find 
\be
\langle \hat{0}|\hat{0}\rangle \ = \ \begin{array}{c}\scalebox{0.6}{
\begin{tikzpicture}[very thick]
\fill[black] (0,0.0) circle (0.05cm);
\fill[black] (0,0.3) circle (0.05cm);
\fill[black] (0,0.6) circle (0.05cm);
\fill[black] (0,0.9) circle (0.05cm);
\draw (0,0) -- (0.4,0) arc (-90:90:0.15cm) -- (0,0.3);
\draw (0,0.6) -- (0.4,0.6) arc (-90:90:0.15cm) -- (0,0.9);
\draw (0,0) -- (-0.4,0) arc (-90:-270:0.15cm) -- (0,0.3);
\draw (0,0.6) -- (-0.4,0.6) arc (-90:-270:0.15cm) -- (0,0.9);
\end{tikzpicture}} 
\end{array} \ = \ d^2\,, \qquad \langle \hat{0}|\hat{1}\rangle \ = \ \begin{array}{c}\scalebox{0.6}{
\begin{tikzpicture}[very thick]
\fill[black] (0,0.0) circle (0.05cm);
\fill[black] (0,0.3) circle (0.05cm);
\fill[black] (0,0.6) circle (0.05cm);
\fill[black] (0,0.9) circle (0.05cm);
\draw (0,0) -- (0.1,0) arc (-90:90:0.45cm) -- (0,0.9);
\draw (0,0.6) -- (0.1,0.6) arc (90:-90:0.15cm) -- (0,0.3);
\draw (0,0) -- (-0.4,0) arc (-90:-270:0.15cm) -- (0,0.3);
\draw (0,0.6) -- (-0.4,0.6) arc (-90:-270:0.15cm) -- (0,0.9);
\end{tikzpicture}} 
\end{array}\ = \ d
\,, \qquad \langle \hat{1}|\hat{1}\rangle \ = \ \begin{array}{c}\scalebox{0.6}{
\begin{tikzpicture}[very thick]
\fill[black] (0,0.0) circle (0.05cm);
\fill[black] (0,0.3) circle (0.05cm);
\fill[black] (0,0.6) circle (0.05cm);
\fill[black] (0,0.9) circle (0.05cm);
\draw (0,0) -- (0.1,0) arc (-90:90:0.45cm) -- (0,0.9);
\draw (0,0.6) -- (0.1,0.6) arc (90:-90:0.15cm) -- (0,0.3);
\draw (0,0) -- (-0.1,0) arc (-90:-270:0.45cm) -- (0,0.9);
\draw (0,0.6) -- (-0.1,0.6) arc (90:270:0.15cm) -- (0,0.3);
\end{tikzpicture}} 
\end{array} \ = \ d^2\,.
\ee
As stated we are not drawing the 3-manifold implying that it is a 3-sphere. So states $|\hat{0}\rangle$ and $|\hat{1}\rangle$ form a nonorthonormal basis in the Hilbert space. Using the skein relation one can shown that all other ways of connecting the punctures can be reduced to a linear combination of~(\ref{CSstates}).

The orthonormal basis can be constructed using the Gram-Schmidt procedure, e.g.
\be
|0\rangle \ = \  \frac{1}{d} \begin{array}{c}\scalebox{0.6}{
\begin{tikzpicture}[very thick]
\fill[black] (0,0.0) circle (0.05cm);
\fill[black] (0,0.3) circle (0.05cm);
\fill[black] (0,0.6) circle (0.05cm);
\fill[black] (0,0.9) circle (0.05cm);
\draw (0,0) -- (0.4,0) arc (-90:90:0.15cm) -- (0,0.3);
\draw (0,0.6) -- (0.4,0.6) arc (-90:90:0.15cm) -- (0,0.9);
\end{tikzpicture}} 
\end{array} \ = \ \frac{1}{d}\,|\hat{0}\rangle\,,\qquad |1\rangle \ = \ \frac{1}{\sqrt{d^2-1}} \left(\begin{array}{c}\scalebox{0.6}{
\begin{tikzpicture}[very thick]
\fill[black] (0,0.0) circle (0.05cm);
\fill[black] (0,0.3) circle (0.05cm);
\fill[black] (0,0.6) circle (0.05cm);
\fill[black] (0,0.9) circle (0.05cm);
\draw (0,0) -- (0.1,0) arc (-90:90:0.45cm) -- (0,0.9);
\draw (0,0.6) -- (0.1,0.6) arc (90:-90:0.15cm) -- (0,0.3);
\end{tikzpicture}} 
\end{array}-\frac{1}{d}\begin{array}{c}\scalebox{0.6}{
\begin{tikzpicture}[very thick]
\fill[black] (0,0.0) circle (0.05cm);
\fill[black] (0,0.3) circle (0.05cm);
\fill[black] (0,0.6) circle (0.05cm);
\fill[black] (0,0.9) circle (0.05cm);
\draw (0,0) -- (0.4,0) arc (-90:90:0.15cm) -- (0,0.3);
\draw (0,0.6) -- (0.4,0.6) arc (-90:90:0.15cm) -- (0,0.9);
\end{tikzpicture}} 
\end{array}\right)\ = \ \frac{1}{\sqrt{d^2-1}} \left(|\hat{1}\rangle-\frac{1}{d}|\hat{0}\rangle\right).
\ee
Note that the orthonormal basis does not have a pure diagrammatic presentation, it is rather given by a linear combination of diagrams, since there is no knot or link whose Jones polynomial vanishes for generic $k$. Note also that for $d=1$ ($k=1$) the norm of the second state is infinite, which means that $|\tilde{0}\rangle$ and $|\tilde{1}\rangle$ are actually linearly dependent. This kind of degeneracy does not happen for $k>1$.

Since there is an explicit diagrammatic presentation for states $|\hat{0}\rangle$ and $|\hat{1}\rangle$, it is convenient to use  the nonorthogonal basis to discuss the topological version of quantum mechanics. A generic state in this quantum mechanics is given by a 3-manifold with a specific way of wiring this manifold with Wilson loops and Wilson lines connecting punctures on the boundary spheres. All overlaps in this basis are the Jones polynomials.

One question that can be asked in the topological picture is whether entanglement equals topology, or rather whether topology can give a good classification of quantum entanglement. This question was addressed in the literature in different ways, e.g.~\cite{Aravind:1997,Kauffman:2013bh,Balasubramanian:2018por,Quinta:2018scm,Melnikov:2022qyt}. In~\cite{Melnikov:2022qyt}, in particular, the present author was using the  setup introduced above to address the problem of the classification. It was shown that for the bipartite entanglement a classification equivalent to the well-known one by Stochastic Local Operations and Classical Communication (SLOCC)~\cite{Dur:2000zz} can be obtained restricting the consideration to the states of simplest topology. Such states were called connectomes and one of the goals of this work is to further clarify the nature of such states.

The simplest topologies discussed in~\cite{Melnikov:2022qyt} are those which mostly contain the information about which party is connected to which, and which connect the parties with a minimal amount of tangling, up to local permutations of punctures. In the bipartite or tripartite case such minimal connectivity can be expressed by planar graphs with two or three vertices respectively. For example for the pair of qubits one has the following three diagrams: 
\be
\label{2qconnectomes}
\begin{array}{c}
\begin{tikzpicture}[thick]
\fill[black] (0,0.0) circle (0.05cm);
\fill[black] (0,0.3) circle (0.05cm);
\fill[black] (0,0.6) circle (0.05cm);
\fill[black] (0,0.9) circle (0.05cm);
\draw (0,0) -- (0.4,0) arc (-90:90:0.15cm) -- (0,0.3);
\draw (0,0.6) -- (0.4,0.6) arc (-90:90:0.15cm) -- (0,0.9);
\fill[black] (1.2,0.0) circle (0.05cm);
\fill[black] (1.2,0.3) circle (0.05cm);
\fill[black] (1.2,0.6) circle (0.05cm);
\fill[black] (1.2,0.9) circle (0.05cm);
\draw (1.2,0) -- (0.8,0) arc (-90:-270:0.15cm) -- (1.2,0.3);
\draw (1.2,0.6) -- (0.8,0.6) arc (-90:-270:0.15cm) -- (1.2,0.9);
\end{tikzpicture} 
\end{array}
\,, \qquad 
\begin{array}{c}
\begin{tikzpicture}[thick]
\fill[black] (0,0.0) circle (0.05cm);
\fill[black] (0,0.3) circle (0.05cm);
\fill[black] (0,0.6) circle (0.05cm);
\fill[black] (0,0.9) circle (0.05cm);
\draw (0,0) -- (0.4,0) -- (1.2,0.0);
\draw (0,0.6) -- (0.4,0.6) arc (-90:90:0.15cm) -- (0,0.9);
\fill[black] (1.2,0.0) circle (0.05cm);
\fill[black] (1.2,0.3) circle (0.05cm);
\fill[black] (1.2,0.6) circle (0.05cm);
\fill[black] (1.2,0.9) circle (0.05cm);
\draw (1.2,0.3) -- (0,0.3);
\draw (1.2,0.6) -- (0.8,0.6) arc (-90:-270:0.15cm) -- (1.2,0.9);
\end{tikzpicture} 
\end{array}
\,,\qquad \text{and}\qquad  
\begin{array}{c}
\begin{tikzpicture}[thick]
\fill[black] (0,0.0) circle (0.05cm);
\fill[black] (0,0.3) circle (0.05cm);
\fill[black] (0,0.6) circle (0.05cm);
\fill[black] (0,0.9) circle (0.05cm);
\draw (0,0) -- (1.2,0);
\draw (0,0.6) -- (1.2,0.6);
\fill[black] (1.2,0.0) circle (0.05cm);
\fill[black] (1.2,0.3) circle (0.05cm);
\fill[black] (1.2,0.6) circle (0.05cm);
\fill[black] (1.2,0.9) circle (0.05cm);
\draw (1.2,0.3) -- (0,0.3);
\draw (1.2,0.9) -- (0,0.9);
\end{tikzpicture} 
\end{array}\,,
\ee
where we show all possible connections of left and right spheres (groups of four points) up to permutations within the individual spheres. The first two diagrams actually represent the same state since two lines are not enough to support the connectivity of left and right~\cite{Melnikov:2022qyt}. Consequently, there are only two independent diagrams, corresponding to separable and Bell class of entanglement. Later we will make a more precise definition of the connectome states observing that in the limit $k\to \infty$ all possible connections reduce to the simplest topologies, similar to~(\ref{2qconnectomes}).

We will also show that the entanglement entropy of the connectome states is given by simple formula, which in the limit of large $k$ and large $n$ reduce to simple counting of connections between the parties. This fact was used in~\cite{Melnikov:2023nzn} to derive connectome versions of the inequalities for the entanglement entropy. Let us consider the example of subadditivity.

\begin{figure}
    \centering
   $ \begin{array}{c}
\begin{tikzpicture}[thick]
\draw (0,0.2) -- (1.5,0.2) -- (1.5,0) -- (0,0) -- (0,-0.2) -- (1.5,-0.2);
\draw (-0.2,0) -- (0.55,-1) -- (0.95,-1) -- (1.7,0);
\fill[white] (0,0) circle (0.3);
\draw (0,0) circle (0.3);
\draw (0,0) node {$A$};
\fill[white] (1.5,0) circle (0.3);
\draw (1.5,0) circle (0.3);
\draw (1.5,0) node {$B$};
\fill[white] (0.75,-1) ellipse (0.6 and 0.3);
\draw (0.75,-1) ellipse (0.6 and 0.3);
\draw (0.75,-1) node {$\overline{AB}$};
\fill[white] (0.75,0) circle (0.21);
\draw[blue] (0.75,0) node {\tiny $\ell_{AB}$};
\draw[blue,dashed,very thick] (0,0.4) arc (90:-120:0.4);
\draw[blue] (-0.15,0.45) node {\tiny $N_A$};
\draw[blue,dashed,very thick] (1.5,0.4) arc (90:300:0.4);
\draw[blue] (1.7,0.45) node {\tiny $N_B$};
\draw[blue,dashed,very thick] (0,-0.8) arc (120:60:1.5);
\draw[blue] (0.75,-0.42) node {\tiny $N_{\overline{AB}}$};
\end{tikzpicture}
\end{array}$
    \caption{Connectome diagram for the subadditivity derivation.}
    \label{fig:subadditivity}
\end{figure}
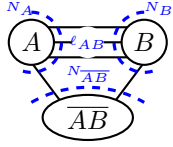

Subadditivity inequality states that the sum of entanglement entropies of subsystems $A$ and $B$ is greater than the entanglement entropy of the union $A\cup B$,
\be
S(A) + S(B) \ \geq \ S(A\cup B)\,.
\ee
A simple connectome derivation can be obtained from the diagram in figure~\ref{fig:subadditivity}. 

Let $\ell_{AB}$ be the number of connections between subsystems $A$ and $B$. Let $N_A$, $N_B$ and $N_{\overline{AB}}$ be the number of connections of subsystems $A$, $B$ and $A\cup B$, respectively, to the rest of the system. Clearly,
\be
N_A + N_B = N_{A\cup B} + 2\ell_{AB} \geq N_{A\cup B}\,,
\ee
which is equivalent to the statement about the entropies of the connectome states in the limit $k\to \infty$, $n\to \infty$, modulo factor $\log 2$, cf. equation~(\ref{dimH}). Using the same counting one can derive many similar inequalities satisfied by entanglement entropies of another special class of quantum states -- the holographic states~\cite{Bao:2015bfa}. To the author's knowledge the connectome states are a subclass of the holographic states, since they satisfy the same inequalities, although in some cases they saturate them.

In the remaining part of this paper we will further explore the connection of the connectomes with the holographic states.

\section{Minimal area formula for entanglement entropy}
\label{sec:entropy}

Let us begin with a discussion of the entanglement entropy of Chern-Simons states with a pair of $S^2$ boundaries.

\subsection{Entanglement entropy}
\label{sec:minarrea}

Let us assume a bipartite state of the form
\be
\label{genconnectome}
\begin{array}{c}
\begin{tikzpicture}[thick]
\newcommand{\x}{1.2}
\fill[cyan] (-0.5,0.55) rectangle (0.2,1.2);
\fill[yellow] (-0.5,-0.1) rectangle (0.2,0.55);
\fill[pink] (\x-0.2,0.55) rectangle (\x+0.5,1.2);
\fill[lime] (\x-0.2,-0.1) rectangle (\x+0.5,0.55);
\foreach \y in {0.6,0.8,...,1.0}
            \draw (0,\y) -- (0.4,\y) arc (-90:90:0.05cm) -- (0,\y+0.1);
\foreach \y in {0.6,0.8,...,1.0}
            \draw (\x,\y) -- (\x-0.4,\y) arc (270:90:0.05cm) -- (\x,\y+0.1);
\foreach \y in {0,0.1,...,0.6}
            \draw (0,\y) -- (\x,\y);
\foreach \y in {0,0.1,...,1.2}
           \fill[black] (0,\y) circle (0.04cm);
\foreach \y in {0,0.1,...,1.2}
           \fill[black] (\x,\y) circle (0.04cm);
\draw (-0.25,0.25) node {$m$};
\draw (-0.25,0.9) node {$n$};
\end{tikzpicture} 
\end{array}\,,
\ee
that is a connectome state. For simplicity, the state is chosen to belong to a product of two equivalent Hilbert spaces $\Hc\otimes\Hc$. There are $m$ connections between punctures in different subsystems and $n$ connection between the punctures of the same subsystem. The density matrix of state~(\ref{genconnectome}) is given by two copies of the above diagram. To construct the reduced density matrix of the left (right) subsystem one simply connects the corresponding punctures of the copies of the right (left) subsystem, as the following diagram shows,
\be
\hat{\rho}_L \ =\ 
\begin{array}{c}
\begin{tikzpicture}[thick]
\newcommand{\x}{1.2}
\newcommand{\z}{0}
\fill[cyan] (-0.5,0.55) rectangle (0.2,1.2);
\fill[yellow] (-0.5,-0.1) rectangle (0.2,0.55);
\fill[pink] (\x-0.2,0.55) rectangle (\x+0.2,1.2);
\fill[lime] (\x-0.2,-0.1) rectangle (\x+0.2,0.55);
\foreach \y in {0.6,0.8,...,1.0}
            \draw (0+\z,\y) -- (0.4+\z,\y) arc (-90:90:0.05cm) -- (0+\z,\y+0.1);
\foreach \y in {0.6,0.8,...,1.0}
            \draw (\x+\z,\y) -- (\x+\z-0.4,\y) arc (270:90:0.05cm) -- (\x+\z,\y+0.1);
\foreach \y in {0,0.1,...,0.6}
            \draw (0+\z,\y) -- (\x+\z,\y);
\foreach \y in {0,0.1,...,1.2}
           \fill[black] (0+\z,\y) circle (0.04cm);
\foreach \y in {0,0.1,...,1.2}
           \fill[black] (\x+\z,\y) circle (0.04cm);
\renewcommand{\z}{1.2}
\fill[cyan] (\x+\z-0.2,0.55) rectangle (\x+\z+0.5,1.2);
\fill[yellow] (\x+\z-0.2,-0.1) rectangle (\x+\z+0.5,0.55);
\foreach \y in {0.6,0.8,...,1.0}
            \draw (0+\z,\y) -- (0.4+\z,\y) arc (-90:90:0.05cm) -- (0+\z,\y+0.1);
\foreach \y in {0.6,0.8,...,1.0}
            \draw (\x+\z,\y) -- (\x+\z-0.4,\y) arc (270:90:0.05cm) -- (\x+\z,\y+0.1);
\foreach \y in {0,0.1,...,0.6}
            \draw (0+\z,\y) -- (\x+\z,\y);
\foreach \y in {0,0.1,...,1.2}
           \fill[black] (\x+\z,\y) circle (0.04cm);
\end{tikzpicture} 
\end{array}
\ = \ d^{n/2} \begin{array}{c}
\begin{tikzpicture}[thick]
\newcommand{\x}{1.2}
\newcommand{\z}{0}
\fill[cyan] (-0.5,0.55) rectangle (0.2,1.2);
\fill[yellow] (-0.5,-0.1) rectangle (0.2,0.55);
\fill[cyan] (\x-0.2,0.55) rectangle (\x+0.5,1.2);
\fill[yellow] (\x-0.2,-0.1) rectangle (\x+0.5,0.55);
\foreach \y in {0.6,0.8,...,1.0}
            \draw (0+\z,\y) -- (0.4+\z,\y) arc (-90:90:0.05cm) -- (0+\z,\y+0.1);
\foreach \y in {0.6,0.8,...,1.0}
            \draw (\x+\z,\y) -- (\x+\z-0.4,\y) arc (270:90:0.05cm) -- (\x+\z,\y+0.1);
\foreach \y in {0,0.1,...,0.6}
            \draw (0+\z,\y) -- (\x+\z,\y);
\foreach \y in {0,0.1,...,1.2}
           \fill[black] (0+\z,\y) circle (0.04cm);
\foreach \y in {0,0.1,...,1.2}
           \fill[black] (\x+\z,\y) circle (0.04cm);
\end{tikzpicture} 
\end{array}\,.
\ee
This is the unnormalized density matrix of the left subsystem. Here we used the rule that every closed line can be substituted by numerical factor $d$.  

In order to compute the trace of $\hat\rho_L$ one should connect the inputs with outputs. Since we consider these lines in the three-dimensional bulk filling the space between two $S^2$, closing produces a copy of $S^2\times S^1$ with $m$ lines winding $S^1$ direction and $n/2$ contractible circles. The contractible circles can again be replaced by $d^{n/2}$ so the trace is given by $m$ non-trivial loops in $S^2\times S^1$.\footnote{For three-dimensional spaces with topology other than $S^3$ a slight modification of rules from section~\ref{sec:basicCS} is necessary. We will only state the results.} The corresponding path integral is just $D_m$ -- the dimension of the Hilbert space of $S^2$ with $m$ punctures~\cite{Witten:1988hf}. Hence,
\be
\label{denmatrix}
\rho_L \ = \ \frac{1}{D_m d^{n/2}} \begin{array}{c}
\begin{tikzpicture}[thick]
\newcommand{\x}{1.2}
\newcommand{\z}{0}
\fill[cyan] (-0.5,0.55) rectangle (0.2,1.2);
\fill[yellow] (-0.5,-0.1) rectangle (0.2,0.55);
\fill[cyan] (\x-0.2,0.55) rectangle (\x+0.5,1.2);
\fill[yellow] (\x-0.2,-0.1) rectangle (\x+0.5,0.55);
\foreach \y in {0.6,0.8,...,1.0}
            \draw (0+\z,\y) -- (0.4+\z,\y) arc (-90:90:0.05cm) -- (0+\z,\y+0.1);
\foreach \y in {0.6,0.8,...,1.0}
            \draw (\x+\z,\y) -- (\x+\z-0.4,\y) arc (270:90:0.05cm) -- (\x+\z,\y+0.1);
\foreach \y in {0,0.1,...,0.6}
            \draw (0+\z,\y) -- (\x+\z,\y);
\foreach \y in {0,0.1,...,1.2}
           \fill[black] (0+\z,\y) circle (0.04cm);
\foreach \y in {0,0.1,...,1.2}
           \fill[black] (\x+\z,\y) circle (0.04cm);
\draw (-0.25,0.25) node {$m$};
\draw (-0.25,0.9) node {$n$};
\end{tikzpicture} 
\end{array}\,, \qquad \text{and} \qquad
\rho^k_L \ = \ \frac{1}{D_m^k d^{n/2}} \begin{array}{c}
\begin{tikzpicture}[thick]
\newcommand{\x}{1.2}
\newcommand{\z}{0}
\fill[cyan] (-0.5,0.55) rectangle (0.2,1.2);
\fill[yellow] (-0.5,-0.1) rectangle (0.2,0.55);
\fill[cyan] (\x-0.2,0.55) rectangle (\x+0.5,1.2);
\fill[yellow] (\x-0.2,-0.1) rectangle (\x+0.5,0.55);
\foreach \y in {0.6,0.8,...,1.0}
            \draw (0+\z,\y) -- (0.4+\z,\y) arc (-90:90:0.05cm) -- (0+\z,\y+0.1);
\foreach \y in {0.6,0.8,...,1.0}
            \draw (\x+\z,\y) -- (\x+\z-0.4,\y) arc (270:90:0.05cm) -- (\x+\z,\y+0.1);
\foreach \y in {0,0.1,...,0.6}
            \draw (0+\z,\y) -- (\x+\z,\y);
\foreach \y in {0,0.1,...,1.2}
           \fill[black] (0+\z,\y) circle (0.04cm);
\foreach \y in {0,0.1,...,1.2}
           \fill[black] (\x+\z,\y) circle (0.04cm);
\draw (-0.25,0.25) node {$m$};
\draw (-0.25,0.9) node {$n$};
\end{tikzpicture} 
\end{array}\,.
\ee
The only thing that distinguishes $\rho_L^k$ from $\rho_L$ is the power of $D_m$. Then
\be
\Tr \rho_L^k \ = \ \frac{1}{D_m^k d^{n/2}}\ \Tr\begin{array}{c}
\begin{tikzpicture}[thick]
\newcommand{\x}{1.2}
\newcommand{\z}{0}
\fill[cyan] (-0.5,0.55) rectangle (0.2,1.2);
\fill[yellow] (-0.5,-0.1) rectangle (0.2,0.55);
\fill[cyan] (\x-0.2,0.55) rectangle (\x+0.5,1.2);
\fill[yellow] (\x-0.2,-0.1) rectangle (\x+0.5,0.55);
\foreach \y in {0.6,0.8,...,1.0}
            \draw (0+\z,\y) -- (0.4+\z,\y) arc (-90:90:0.05cm) -- (0+\z,\y+0.1);
\foreach \y in {0.6,0.8,...,1.0}
            \draw (\x+\z,\y) -- (\x+\z-0.4,\y) arc (270:90:0.05cm) -- (\x+\z,\y+0.1);
\foreach \y in {0,0.1,...,0.6}
            \draw (0+\z,\y) -- (\x+\z,\y);
\foreach \y in {0,0.1,...,1.2}
           \fill[black] (0+\z,\y) circle (0.04cm);
\foreach \y in {0,0.1,...,1.2}
           \fill[black] (\x+\z,\y) circle (0.04cm);
\draw (-0.25,0.25) node {$m$};
\draw (-0.25,0.9) node {$n$};
\end{tikzpicture} 
\end{array} \ = \ \frac{1}{D_m^k}\ \Tr\begin{array}{c}
\begin{tikzpicture}[thick]
\newcommand{\x}{1.2}
\newcommand{\z}{0}
\fill[yellow] (-0.5,-0.1) rectangle (0.2,0.55);
\fill[yellow] (\x-0.2,-0.1) rectangle (\x+0.5,0.55);
\foreach \y in {0,0.1,...,0.6}
            \draw (0+\z,\y) -- (\x+\z,\y);
\foreach \y in {0,0.1,...,0.6}
           \fill[black] (0+\z,\y) circle (0.04cm);
\foreach \y in {0,0.1,...,0.6}
           \fill[black] (\x+\z,\y) circle (0.04cm);
\draw (-0.25,0.25) node {$m$};
\end{tikzpicture} 
\end{array} \ = \ \frac{1}{D^{k-1}_m}\,.
\ee
Computing the entropy yields
\be
S \ = \ \log D_m\,.
\ee
In other words, the entanglement entropy is defined by the connected part of the diagram, and more precisely, by the dimension of the subspace of the Hilbert space associated with the punctures connected to the other subsystem.

Note that we would obtain the same result for any state that can be obtained from~(\ref{genconnectome}) by local unitary transformations, for example permutation of the lines by the left or right action of unitary braiding operators. This follows from the invariance of the definition of the von Neumann entropy under local unitary transformations.

One can view this result in the following way. The diagrams of the reduced density matrix in formula~(\ref{denmatrix}) illustrate an example of two spheres connected by lines. One can cut the bulk separating the two spheres by placing another surface of spherical topology between the two (by enclosing one of the spheres). This surface will have a certain number of intersections with the lines. It is possible to choose the surface in such a way that it will be crossed by the minimal number of lines. The entanglement entropy is defined by this minimal number. 

In the classical limit of $SU(2)_k$ Chern-Simons theory, $k\to\infty$, and large $m\ll k$, dimension $D_m$ of the Hilbert space of the sphere with $m$ punctures is given approximately by
\be
D_m \ \simeq \  C_{m/2} \ \simeq \ \frac{4^{m/2}}{(m/2)^{3/2}\sqrt{\pi}}\,,
\ee
so that the entanglement entropy is approximately $m\log 2$.

The above result for the entanglement entropy is an analog of the Ryu-Takayanagi formula for the holographic states~\cite{Ryu:2006bv}. The formula claims that for theories with a gravity dual the entanglement entropy of a $d$-dimensional region $A$ is given by the area of a surface $\gamma_A$ that is homological to $A$ in the $(d+1)$-dimensional bulk space and that has minimal area,
\be
S \ = \ \min\limits_{\gamma_A}\frac{{\rm Area}[\gamma_A]}{4G}\,,
\ee
where $G$ is the Newton's constant. The Ryu-Takayanagi formula is valid in the classical limit of the holographic correspondence, that is when the areas are very large in Planck units, $G\to 0$. If one thinks of Wilson lines as flux lines carrying a unit of flux, then the entropy is indeed proportional to the area divided by a typical area per flux, which is of the order of $\hbar^2$.

The discrete version of the Ryu-Takayanagi formula appear in the tensor network constructions of holography~\cite{Pastawski:2015qua}. Chern-Simons theory is of course a special example of a tensor network. Moreover, the Chern-Simons example is a version of the bit thread approach to the calculation of the entanglement entropy discussed in~\cite{Freedman:2016zud}. The example considered here points to a special class of states for which the Ryu-Takayanagi formula is valid -- these are the connectome states. Let us discuss examples of other states and see how the minimal area formula is modified.

\subsection{Corrections to the minimal area formula}
\label{sec:corrections}

\begin{figure}
    \centering
    \includegraphics[width=0.5\linewidth]{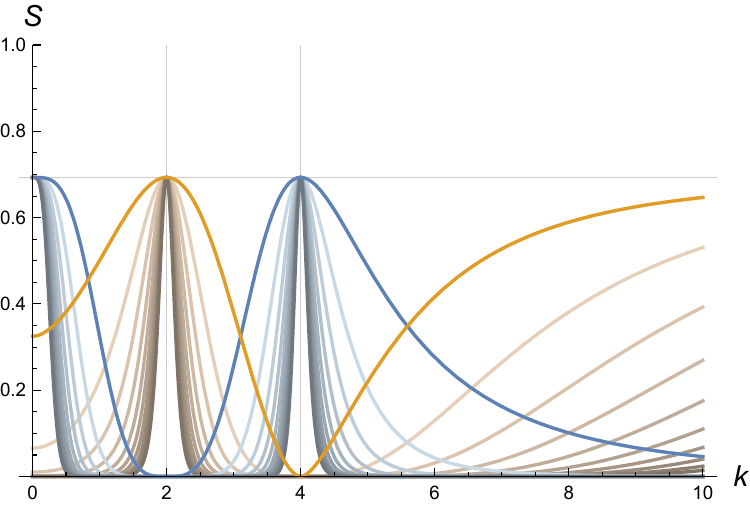}
    \caption{Entantglement entropy of states~(\ref{chainedstate0}) (bright blue) and~(\ref{ringedstate}) (bright orange) as a function of $k$. Shaded colors show the entropy of the subfamilies~(\ref{subfamily}) of these states, with color darkness increasing with $n$, leaving only sharp peaks at $k=0,2,4,\infty$ in the limit $n\to\infty$. The horizontal line indicates the maximum value $\log2$. }
    \label{fig:chainentropy}
\end{figure}

For general quantum states the entanglement entropy is not calculated by the above minimal area formula. In particular, the formula is not valid for linear combinations of the connectome states. As an example we can consider state
\begin{multline}
\label{chainedstate0}
\begin{array}{c}
\begin{tikzpicture}[thick]
\newcommand{\x}{0.6}
\draw (0,0) -- (1.8,0);
\draw (0,0.3) -- (1.8,0.3);
\draw (0,0.6) -- (0.5,0.6) arc (-90:30:0.15cm);
\draw (0,0.9) -- (0.5,0.9) arc (90:60:0.15cm);
\draw (0.9,0.6) -- (0.7,0.6) arc (-90:-120:0.15cm);
\draw (0.9,0.6) -- (1.1,0.6) arc (-90:-60:0.15cm);
\draw (0.9,0.9) -- (0.7,0.9) arc (90:210:0.15cm);
\draw (0.9,0.9) -- (1.1,0.9) arc (90:-30:0.15cm);
\draw (1.2+\x,0.9) -- (0.7+\x,0.9) arc (90:120:0.15cm);
\draw (1.2+\x,0.6) -- (0.7+\x,0.6) arc (-90:-210:0.15cm);
\fill[black] (1.2+\x,0.0) circle (0.05cm);
\fill[black] (1.2+\x,0.3) circle (0.05cm);
\fill[black] (1.2+\x,0.6) circle (0.05cm);
\fill[black] (1.2+\x,0.9) circle (0.05cm);
\fill[black] (0,0.0) circle (0.05cm);
\fill[black] (0,0.3) circle (0.05cm);
\fill[black] (0,0.6) circle (0.05cm);
\fill[black] (0,0.9) circle (0.05cm);
\end{tikzpicture} 
\end{array}
\ = \ (-A^6-A^{-6}) 
\begin{array}{c}
\begin{tikzpicture}[thick]
\fill[black] (0,0.0) circle (0.05cm);
\fill[black] (0,0.3) circle (0.05cm);
\fill[black] (0,0.6) circle (0.05cm);
\fill[black] (0,0.9) circle (0.05cm);
\draw (0,0) -- (0.4,0) -- (1.2,0.0);
\draw (0,0.6) -- (0.4,0.6) arc (-90:90:0.15cm) -- (0,0.9);
\fill[black] (1.2,0.0) circle (0.05cm);
\fill[black] (1.2,0.3) circle (0.05cm);
\fill[black] (1.2,0.6) circle (0.05cm);
\fill[black] (1.2,0.9) circle (0.05cm);
\draw (1.2,0.3) -- (0,0.3);
\draw (1.2,0.6) -- (0.8,0.6) arc (-90:-270:0.15cm) -- (1.2,0.9);
\end{tikzpicture} 
\end{array}
 +  (1-A^{-4})(1-A^{4})  
\begin{array}{c}
\begin{tikzpicture}[thick]
\draw (0,0) -- (1.2,0);
\draw (0,0.6) -- (1.2,0.6);
\draw (1.2,0.3) -- (0,0.3);
\draw (1.2,0.9) -- (0,0.9);
\fill[black] (0,0.0) circle (0.05cm);
\fill[black] (0,0.3) circle (0.05cm);
\fill[black] (0,0.6) circle (0.05cm);
\fill[black] (0,0.9) circle (0.05cm);
\fill[black] (1.2,0.0) circle (0.05cm);
\fill[black] (1.2,0.3) circle (0.05cm);
\fill[black] (1.2,0.6) circle (0.05cm);
\fill[black] (1.2,0.9) circle (0.05cm);
\end{tikzpicture} 
\end{array}
\\ = \ (A^4+A^{-4})^2|00\rangle + (1-A^{-4})(1-A^{4})|11\rangle\,.
\end{multline}
The entanglement entropy of this state is given by
\be
\label{entent}
S \ = \ - \frac{|s|^2}{|s|^2+|c|^2}\log\frac{|s|^2}{|s|^2+|c|^2} - \frac{|c|^2}{|s|^2+|c|^2}\log \frac{|s|^2}{|s|^2+|c|^2}\,,
\ee
where
\be
|c| \ = \ 4\,\cos^2\!\left(\frac{2\pi}{k+2}\right)\,, \qquad |s| \ = \  4\,\sin^2\!\left(\frac{\pi}{k+2}\right)\,.
\ee
The entropy as a function of $k$ is shown in figure~\ref{fig:chainentropy} (saturated blue). Except for $k=0$ and $k=4$ the entropy is lower than the maximal value $\log 2$. It vanishes in the $k\to\infty$ limit. The minimal area law gives an upper bound and the lower bound on the entropy, since the state is a linear combination of two states with entropies defined by the minimal area law. We note that it is not even clear how to apply the minimal area law due to the presence of crossings in the diagram.

Another useful and connected example is the state
\begin{multline}
\label{ringedstate}
\begin{array}{c}
\begin{tikzpicture}[thick]
\draw[xscale=0.5] (1.2,0.45) arc (270:90:0.3);
\draw[line width=0.1cm,white] (0,0.6) -- (1.2,0.6);
\draw[line width=0.1cm,white] (0,0.9) -- (1.2,0.9);
\draw (0,0) -- (1.2,0);
\draw (0,0.6) -- (1.2,0.6);
\draw (1.2,0.3) -- (0,0.3);
\draw (1.2,0.9) -- (0,0.9);
\draw[line width=0.1cm,white,xscale=0.5] (1.2,0.45) arc (-90:80:0.3);
\draw[xscale=0.5] (1.2,0.45) arc (-90:90:0.3);
\fill[black] (0,0.0) circle (0.05cm);
\fill[black] (0,0.3) circle (0.05cm);
\fill[black] (0,0.6) circle (0.05cm);
\fill[black] (0,0.9) circle (0.05cm);
\fill[black] (1.2,0.0) circle (0.05cm);
\fill[black] (1.2,0.3) circle (0.05cm);
\fill[black] (1.2,0.6) circle (0.05cm);
\fill[black] (1.2,0.9) circle (0.05cm);
\end{tikzpicture} 
\end{array}
=  (1-A^{-4})(1-A^4)
\begin{array}{c}
\begin{tikzpicture}[thick]
\fill[black] (0,0.0) circle (0.05cm);
\fill[black] (0,0.3) circle (0.05cm);
\fill[black] (0,0.6) circle (0.05cm);
\fill[black] (0,0.9) circle (0.05cm);
\draw (0,0) -- (0.4,0) -- (1.2,0.0);
\draw (0,0.6) -- (0.4,0.6) arc (-90:90:0.15cm) -- (0,0.9);
\fill[black] (1.2,0.0) circle (0.05cm);
\fill[black] (1.2,0.3) circle (0.05cm);
\fill[black] (1.2,0.6) circle (0.05cm);
\fill[black] (1.2,0.9) circle (0.05cm);
\draw (1.2,0.3) -- (0,0.3);
\draw (1.2,0.6) -- (0.8,0.6) arc (-90:-270:0.15cm) -- (1.2,0.9);
\end{tikzpicture} 
\end{array}
 +   (-A^6-A^{-6})  
\begin{array}{c}
\begin{tikzpicture}[thick]
\draw (0,0) -- (1.2,0);
\draw (0,0.6) -- (1.2,0.6);
\draw (1.2,0.3) -- (0,0.3);
\draw (1.2,0.9) -- (0,0.9);
\fill[black] (0,0.0) circle (0.05cm);
\fill[black] (0,0.3) circle (0.05cm);
\fill[black] (0,0.6) circle (0.05cm);
\fill[black] (0,0.9) circle (0.05cm);
\fill[black] (1.2,0.0) circle (0.05cm);
\fill[black] (1.2,0.3) circle (0.05cm);
\fill[black] (1.2,0.6) circle (0.05cm);
\fill[black] (1.2,0.9) circle (0.05cm);
\end{tikzpicture} 
\end{array} \\ = \ (-A^2-A^{-2})|00\rangle + (-A^6-A^{-6}) |11\rangle.
\end{multline}
Note that the expansion in terms of the connectome diagrams in the first line gives the swapped coefficients as compared to~(\ref{chainedstate0}). This is the consequence of the fact that the original diagrams are related by a non-local permutation of the upper pairs of points and serves as an illustration of convenient features of the topological presentation. For~(\ref{ringedstate}) the entanglement entropy is given by equation~(\ref{entent}) with
\be
|s| \ = \ 2\,\cos\!\left(\frac{\pi}{k+2}\right)\,, \qquad |c| \ = \ 2\,\cos\!\left(\frac{3\pi}{k+2}\right)\,.
\ee
This entropy is shown in figure~\ref{fig:chainentropy} (saturated orange). It is zero for $k=4$ and has maximal value $\log 2$ for $k=2$ and $k\to\infty$.

It is straightforward to generalize the above two examples to the case of states with multiple loops, obtained by concatenation of the diagrams~(\ref{chainedstate0}) and~(\ref{ringedstate}), that is
\be
\label{subfamily}
\left(\begin{array}{c}
\begin{tikzpicture}[thick]
\newcommand{\x}{0.6}
\draw (0,0) -- (1.8,0);
\draw (0,0.3) -- (1.8,0.3);
\draw (0,0.6) -- (0.5,0.6) arc (-90:30:0.15cm);
\draw (0,0.9) -- (0.5,0.9) arc (90:60:0.15cm);
\draw (0.9,0.6) -- (0.7,0.6) arc (-90:-120:0.15cm);
\draw (0.9,0.6) -- (1.1,0.6) arc (-90:-60:0.15cm);
\draw (0.9,0.9) -- (0.7,0.9) arc (90:210:0.15cm);
\draw (0.9,0.9) -- (1.1,0.9) arc (90:-30:0.15cm);
\draw (1.2+\x,0.9) -- (0.7+\x,0.9) arc (90:120:0.15cm);
\draw (1.2+\x,0.6) -- (0.7+\x,0.6) arc (-90:-210:0.15cm);
\fill[black] (1.2+\x,0.0) circle (0.05cm);
\fill[black] (1.2+\x,0.3) circle (0.05cm);
\fill[black] (1.2+\x,0.6) circle (0.05cm);
\fill[black] (1.2+\x,0.9) circle (0.05cm);
\fill[black] (0,0.0) circle (0.05cm);
\fill[black] (0,0.3) circle (0.05cm);
\fill[black] (0,0.6) circle (0.05cm);
\fill[black] (0,0.9) circle (0.05cm);
\end{tikzpicture} 
\end{array}\right)^n\, \qquad \text{and} \qquad \left(\begin{array}{c}
\begin{tikzpicture}[thick]
\draw[xscale=0.5] (1.2,0.45) arc (270:90:0.3);
\draw[line width=0.1cm,white] (0,0.6) -- (1.2,0.6);
\draw[line width=0.1cm,white] (0,0.9) -- (1.2,0.9);
\draw (0,0) -- (1.2,0);
\draw (0,0.6) -- (1.2,0.6);
\draw (1.2,0.3) -- (0,0.3);
\draw (1.2,0.9) -- (0,0.9);
\draw[line width=0.1cm,white,xscale=0.5] (1.2,0.45) arc (-90:80:0.3);
\draw[xscale=0.5] (1.2,0.45) arc (-90:90:0.3);
\fill[black] (0,0.0) circle (0.05cm);
\fill[black] (0,0.3) circle (0.05cm);
\fill[black] (0,0.6) circle (0.05cm);
\fill[black] (0,0.9) circle (0.05cm);
\fill[black] (1.2,0.0) circle (0.05cm);
\fill[black] (1.2,0.3) circle (0.05cm);
\fill[black] (1.2,0.6) circle (0.05cm);
\fill[black] (1.2,0.9) circle (0.05cm);
\end{tikzpicture} 
\end{array}\right)^n\,.
\ee
With the increase of the number of loops, or equivalently, with the increase of the complexity of the diagram, the entanglement between the parties decreases, as evidenced by the shaded curves in figure~\ref{fig:chainentropy}. The presence of many loops decreases the connectivity of the space, making it disconnected eventually, and therefore unentangled. 

A qualitatively different class of states that appear in Chern-Simons theory features distinct three-dimensional topologies. Even if we fix the boundaries to be $S^2$, the bulk is not necessarily globally $S^3$. Let us consider an example with lines in $S^2\times S^1$ with a pair of $S^2$ boundaries.
\be
\label{holedstate}
\begin{array}{c}
    \begin{tikzpicture}[thick,xscale=cos(70)]
    \draw[double distance=1.4mm,opacity=0.5] (0:1) arc (-90:-270:0.3);
    \draw[rounded corners = 2] (-2,0.15) -- (-1.1,0.15) -- (-1.1,0.35) -- (3.1,0.35) -- (3.1,0.15) -- (4,0.15);
    \draw (-2,0.45) -- (4,0.45);
    \foreach \y in {0,0.1,...,0.2}
           \draw (0,0.15+\y) -- (2,0.15+\y);
    \draw[double distance=1.4mm,opacity=0.7] (0:1) arc (-90:90:0.3);
    \draw[rounded corners = 2] (2,0.15) -- (2.5,0.15) -- (2.5,-0.15) -- (-0.5,-0.15) -- (-0.5,0.15) -- (0,0.15);
    \draw[rounded corners = 2] (2,0.25) -- (2.8,0.25) -- (2.8,-0.25) -- (-0.8,-0.25) -- (-0.8,0.25) -- (0,0.25);
    \draw[rounded corners = 2] (-2,-0.15) -- (-1.1,-0.15) -- (-1.1,-0.35) -- (3.1,-0.35) -- (3.1,-0.15) -- (4,-0.15);
    \draw (-2,-0.45) -- (4,-0.45);
    \foreach \y in {0,0.3,...,1.2}
           \fill[black] (-2,-0.45+\y) ellipse (0.15cm and 0.05cm);
    \foreach \y in {0,0.3,...,1.2}
           \fill[black] (4,-0.45+\y) ellipse (0.15cm and 0.05cm);
  \end{tikzpicture}
\end{array} \ = \ |00\rangle  + \frac{1}{(d^2-1)}|11\rangle \ = \ \frac{d^2-2}{d(d^2-1)}\begin{array}{c}
\begin{tikzpicture}[thick]
\fill[black] (0,0.0) circle (0.05cm);
\fill[black] (0,0.3) circle (0.05cm);
\fill[black] (0,0.6) circle (0.05cm);
\fill[black] (0,0.9) circle (0.05cm);
\draw (0,0) -- (0.4,0) -- (1.2,0.0);
\draw (0,0.6) -- (0.4,0.6) arc (-90:90:0.15cm) -- (0,0.9);
\fill[black] (1.2,0.0) circle (0.05cm);
\fill[black] (1.2,0.3) circle (0.05cm);
\fill[black] (1.2,0.6) circle (0.05cm);
\fill[black] (1.2,0.9) circle (0.05cm);
\draw (1.2,0.3) -- (0,0.3);
\draw (1.2,0.6) -- (0.8,0.6) arc (-90:-270:0.15cm) -- (1.2,0.9);
\end{tikzpicture} 
\end{array}
 +    \frac{1}{(d^2-1)} 
\begin{array}{c}
\begin{tikzpicture}[thick]
\draw (0,0) -- (1.2,0);
\draw (0,0.6) -- (1.2,0.6);
\draw (1.2,0.3) -- (0,0.3);
\draw (1.2,0.9) -- (0,0.9);
\fill[black] (0,0.0) circle (0.05cm);
\fill[black] (0,0.3) circle (0.05cm);
\fill[black] (0,0.6) circle (0.05cm);
\fill[black] (0,0.9) circle (0.05cm);
\fill[black] (1.2,0.0) circle (0.05cm);
\fill[black] (1.2,0.3) circle (0.05cm);
\fill[black] (1.2,0.6) circle (0.05cm);
\fill[black] (1.2,0.9) circle (0.05cm);
\end{tikzpicture} 
\end{array} \,.
\ee
Here the torus represents the defect region, where the space differs from $S^3$. For the space to be indeed $S^2\times S^1$, the interior of the torus should be contractible along what appears as a non-contractible cycle and non-contractible along the contractible cycle. In other words one should fill the interior with a solid torus whose fundamental cycles are exchanged. This is how $S^2\times S^1$ is obtained from $S^3$ via surgery operation~\cite{Witten:1988hf}.

The entropy of state~(\ref{holedstate}) is shown in figure (saturated blue). It is only maximal for $k=0$ and $k=2$. For $k\to\infty$ the entropy asymptotes to
\be
\label{topasympt}
S \ = \ \log\frac{10}{3^{9/5}}\ \simeq \ 0.325\,.
\ee

\begin{figure}
    \centering
    \includegraphics[width=0.5\linewidth]{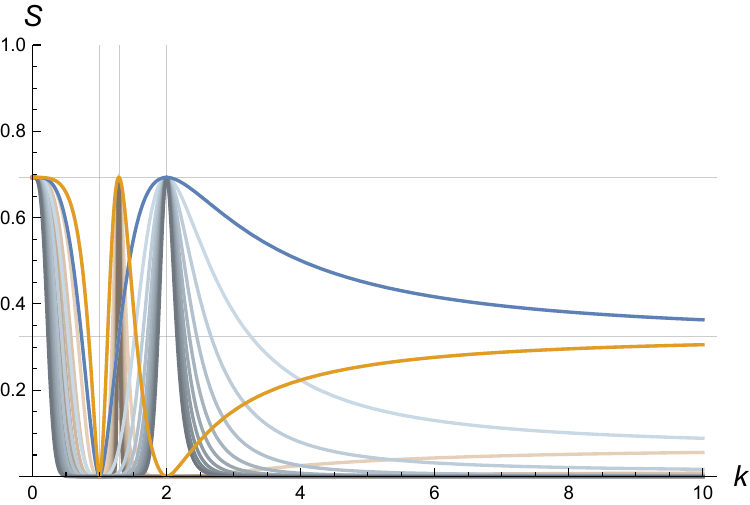}
    \caption{Entantglement entropy of states~(\ref{holedstate}) (saturated blue) and~(\ref{holedstate1}) (saturated orange) as functions of $k$. Shaded colors show the case of the analogs of subfamilies~(\ref{subfamily}), as in figure~\ref{fig:chainentropy}. The horizontal grids show the maximal entropy $\log 2$ and the value~(\ref{topasympt}).}
    \label{fig:holedentropy}
\end{figure}

Let us consider one more example
\be
\label{holedstate1}
\begin{array}{c}
    \begin{tikzpicture}[thick]
    \draw[double distance=3mm,opacity=0.5,yscale=cos(50)] (0:1) arc (180:0:0.4);
    \draw[rounded corners=2] (0.8,0.15) -- (1.25,0.15) -- (1.25,-0.15) -- (0.7,-0.15) -- (0.7,0.15) -- (0.8,0.15);
    \draw[rounded corners=2] (2.0,0.15) -- (1.55,0.15) -- (1.55,-0.15) -- (2.1,-0.15) -- (2.1,0.15) -- (2.0,0.15);
    \draw[rounded corners = 2] (0.2,0.) -- (0.6,0.) -- (0.6,-0.3) -- (1.35,-0.3) -- (1.35,0.3) -- (0.2,0.3);
    \draw[rounded corners = 2] (2.6,0.) -- (2.2,0.) -- (2.2,-0.3) -- (1.45,-0.3) -- (1.45,0.3) -- (2.6,0.3);
    \draw[double distance=3mm,opacity=0.7,yscale=cos(50)] (0:1) arc (-180:0:0.4);
    \draw[rounded corners = 2] (0.2,-0.3) -- (0.5,-0.3) -- (0.5,-0.5) -- (2.3,-0.5) -- (2.3,-0.3) -- (2.6,-0.3);
    \draw (0.2,-0.6) -- (2.6,-0.6);
    \foreach \y in {0,0.3,...,1.2}
           \fill[black] (0.2,-0.6+\y) circle (0.05cm);
    \foreach \y in {0,0.3,...,1.2}
           \fill[black] (2.6,-0.6+\y) circle (0.05cm);
  \end{tikzpicture}
\end{array} \ =\  \frac{1}{(d^2-1)}\begin{array}{c}
\begin{tikzpicture}[thick]
\fill[black] (0,0.0) circle (0.05cm);
\fill[black] (0,0.3) circle (0.05cm);
\fill[black] (0,0.6) circle (0.05cm);
\fill[black] (0,0.9) circle (0.05cm);
\draw (0,0) -- (0.4,0) -- (1.2,0.0);
\draw (0,0.6) -- (0.4,0.6) arc (-90:90:0.15cm) -- (0,0.9);
\fill[black] (1.2,0.0) circle (0.05cm);
\fill[black] (1.2,0.3) circle (0.05cm);
\fill[black] (1.2,0.6) circle (0.05cm);
\fill[black] (1.2,0.9) circle (0.05cm);
\draw (1.2,0.3) -- (0,0.3);
\draw (1.2,0.6) -- (0.8,0.6) arc (-90:-270:0.15cm) -- (1.2,0.9);
\end{tikzpicture} 
\end{array}
 +   \frac{d^2-2}{d(d^2-1)}
\begin{array}{c}
\begin{tikzpicture}[thick]
\draw (0,0) -- (1.2,0);
\draw (0,0.6) -- (1.2,0.6);
\draw (1.2,0.3) -- (0,0.3);
\draw (1.2,0.9) -- (0,0.9);
\fill[black] (0,0.0) circle (0.05cm);
\fill[black] (0,0.3) circle (0.05cm);
\fill[black] (0,0.6) circle (0.05cm);
\fill[black] (0,0.9) circle (0.05cm);
\fill[black] (1.2,0.0) circle (0.05cm);
\fill[black] (1.2,0.3) circle (0.05cm);
\fill[black] (1.2,0.6) circle (0.05cm);
\fill[black] (1.2,0.9) circle (0.05cm);
\end{tikzpicture} 
\end{array} \ = \ \frac{2}{d}\,|00\rangle  + \frac{d^2-2}{d(d^2-1)}|11\rangle\,,
\ee
where we applied the same trick with reshuffling of the lines in order to use the available coefficients of equation~(\ref{holedstate}). The entropy of the last state is shown in figure~\ref{fig:holedentropy} (saturated orange). It has zero value at $k=1,2$ and the maximum value for $k=0$ and for non-integer $k\simeq 1.3$. For $k\to \infty$ it approaches the value given by equation~(\ref{topasympt}). As for states~(\ref{subfamily}), presence of multiple three-dimensional defects decreases the space connectivity and, consequently, the entropy, as illustrated by the family of shaded curves in figure~\ref{fig:holedentropy}.

\section{Classical limit}
\label{sec:classlimit}

The limit $k\to\infty$ is the classical limit of Chern-Simons theory. In this limit  $A\to 1$ ($q\to 1$) and braiding becomes a simple permutation,
\be
\begin{array}{c}
\begin{tikzpicture}[baseline=0]
\draw[thick,rounded corners=2] (0.5,-0.1) -- (0.3,-0.1) -- (0.15,0);
\draw[thick,rounded corners=2] (-0.15,0.2) -- (-0.3,0.3) -- (-0.5,0.3);
\draw[thick,rounded corners=2] (-0.5,-0.1) -- (-0.3,-0.1) -- (0.3,0.3) -- (0.5,0.3);
\end{tikzpicture}
\end{array} \ = \ \begin{array}{c}
\begin{tikzpicture}[baseline=0]
\draw[thick] (-0.5,-0.1) -- (-0.3,-0.1) arc (-90:90:0.2) -- (-0.5,0.3);
\draw[thick] (0.5,-0.1) -- (0.3,-0.1) arc (-90:-270:0.2) -- (0.5,0.3);
\end{tikzpicture}
\end{array} + \begin{array}{c}
\begin{tikzpicture}[baseline=0]
\draw[thick] (0.5,-0.1) -- (-0.5,-0.1);
\draw[thick] (0.5,0.3) -- (-0.5,0.3);
\end{tikzpicture}
\end{array}\,,
\qquad \text{hence} \qquad 
\begin{array}{c}
\begin{tikzpicture}[baseline=0]
\draw[thick,rounded corners=2] (0.5,-0.1) -- (0.3,-0.1) -- (0.15,0);
\draw[thick,rounded corners=2] (-0.15,0.2) -- (-0.3,0.3) -- (-0.5,0.3);
\draw[thick,rounded corners=2] (-0.5,-0.1) -- (-0.3,-0.1) -- (0.3,0.3) -- (0.5,0.3);
\end{tikzpicture}
\end{array} \ = \ \begin{array}{c}
\begin{tikzpicture}[baseline=0,xscale=-1]
\draw[thick,rounded corners=2] (0.5,-0.1) -- (0.3,-0.1) -- (0.15,0);
\draw[thick,rounded corners=2] (-0.15,0.2) -- (-0.3,0.3) -- (-0.5,0.3);
\draw[thick,rounded corners=2] (-0.5,-0.1) -- (-0.3,-0.1) -- (0.3,0.3) -- (0.5,0.3);
\end{tikzpicture}
\end{array}\,.
\ee
(To prove this just rotate 90 degrees all the diagrams in the skein relation.) The latter property means that lines can pass through each other and knots become trivial in the classical limit. Consequently, the states represented by diagrams, like~(\ref{chainedstate0}) and~(\ref{ringedstate}), must all reduce to states with trivial topological linking, that is connectome states. In the above examples,
\be
\begin{array}{c}
\begin{tikzpicture}[thick]
\newcommand{\x}{0.6}
\draw (0,0) -- (1.8,0);
\draw (0,0.3) -- (1.8,0.3);
\draw (0,0.6) -- (0.5,0.6) arc (-90:30:0.15cm);
\draw (0,0.9) -- (0.5,0.9) arc (90:60:0.15cm);
\draw (0.9,0.6) -- (0.7,0.6) arc (-90:-120:0.15cm);
\draw (0.9,0.6) -- (1.1,0.6) arc (-90:-60:0.15cm);
\draw (0.9,0.9) -- (0.7,0.9) arc (90:210:0.15cm);
\draw (0.9,0.9) -- (1.1,0.9) arc (90:-30:0.15cm);
\draw (1.2+\x,0.9) -- (0.7+\x,0.9) arc (90:120:0.15cm);
\draw (1.2+\x,0.6) -- (0.7+\x,0.6) arc (-90:-210:0.15cm);
\fill[black] (1.2+\x,0.0) circle (0.05cm);
\fill[black] (1.2+\x,0.3) circle (0.05cm);
\fill[black] (1.2+\x,0.6) circle (0.05cm);
\fill[black] (1.2+\x,0.9) circle (0.05cm);
\fill[black] (0,0.0) circle (0.05cm);
\fill[black] (0,0.3) circle (0.05cm);
\fill[black] (0,0.6) circle (0.05cm);
\fill[black] (0,0.9) circle (0.05cm);
\end{tikzpicture} 
\end{array} \ \to \ \begin{array}{c}
\begin{tikzpicture}[thick]
\fill[black] (0,0.0) circle (0.05cm);
\fill[black] (0,0.3) circle (0.05cm);
\fill[black] (0,0.6) circle (0.05cm);
\fill[black] (0,0.9) circle (0.05cm);
\draw (0,0) -- (0.4,0) -- (1.2,0.0);
\draw (0,0.6) -- (0.4,0.6) arc (-90:90:0.15cm) -- (0,0.9);
\fill[black] (1.2,0.0) circle (0.05cm);
\fill[black] (1.2,0.3) circle (0.05cm);
\fill[black] (1.2,0.6) circle (0.05cm);
\fill[black] (1.2,0.9) circle (0.05cm);
\draw (1.2,0.3) -- (0,0.3);
\draw (1.2,0.6) -- (0.8,0.6) arc (-90:-270:0.15cm) -- (1.2,0.9);
\end{tikzpicture} 
\end{array}\,,
\qquad \text{and} \qquad
\begin{array}{c}
\begin{tikzpicture}[thick]
\draw[xscale=0.5] (1.2,0.45) arc (270:90:0.3);
\draw[line width=0.1cm,white] (0,0.6) -- (1.2,0.6);
\draw[line width=0.1cm,white] (0,0.9) -- (1.2,0.9);
\draw (0,0) -- (1.2,0);
\draw (0,0.6) -- (1.2,0.6);
\draw (1.2,0.3) -- (0,0.3);
\draw (1.2,0.9) -- (0,0.9);
\draw[line width=0.1cm,white,xscale=0.5] (1.2,0.45) arc (-90:80:0.3);
\draw[xscale=0.5] (1.2,0.45) arc (-90:90:0.3);
\fill[black] (0,0.0) circle (0.05cm);
\fill[black] (0,0.3) circle (0.05cm);
\fill[black] (0,0.6) circle (0.05cm);
\fill[black] (0,0.9) circle (0.05cm);
\fill[black] (1.2,0.0) circle (0.05cm);
\fill[black] (1.2,0.3) circle (0.05cm);
\fill[black] (1.2,0.6) circle (0.05cm);
\fill[black] (1.2,0.9) circle (0.05cm);
\end{tikzpicture} 
\end{array}
\ \to \ 
\begin{array}{c}
\begin{tikzpicture}[thick]
\draw (0,0) -- (1.2,0);
\draw (0,0.6) -- (1.2,0.6);
\draw (1.2,0.3) -- (0,0.3);
\draw (1.2,0.9) -- (0,0.9);
\fill[black] (0,0.0) circle (0.05cm);
\fill[black] (0,0.3) circle (0.05cm);
\fill[black] (0,0.6) circle (0.05cm);
\fill[black] (0,0.9) circle (0.05cm);
\fill[black] (1.2,0.0) circle (0.05cm);
\fill[black] (1.2,0.3) circle (0.05cm);
\fill[black] (1.2,0.6) circle (0.05cm);
\fill[black] (1.2,0.9) circle (0.05cm);
\end{tikzpicture} 
\end{array}\,,
\ee
which explains the asymptotic behavior of the entanglement entropy in figure~\ref{fig:chainentropy}. Consequently, classical limit of Chern-Simons gives precise definition of the connectome states as the classical states in the global $S^3$ topology up to local permutations of the endpoints on the boundaries. In quantum theory connectomes label equivalence classes of states that have the same classical limit.

Since in the classical limit all the states become connectomes the entanglement entropy  is given by the minimal area law for all of them, as long as we consider the situation of bipartition by two boundary spheres.\footnote{More general situations will be discussed in the next section.} Away from the classical regime the minimal area law must be corrected. Since all the observables in Chern-Simons theory are computable, one can study the corrections at any order. For state~(\ref{chainedstate0}) the leading corrections to the entropy are given by
\be
S \ = \ \frac{\pi^4}{k^4}\left(1-\log\frac{\pi^4}{k^4}\right) + O(k^{-5})\,.
\ee
As we have already seen, the entropy decays in the classical limit, since the asymptotic state is separable.

Similarly, for state~(\ref{ringedstate}) the asymptotic expansion of the entropy is 
\be
S \ = \ \log 2 - \frac{8\pi^4}{k^4} +  O(k^{-5})\,,
\ee
and the state is asymptotically maximally entangled.

Comparing these findings with the AdS/CFT correspondence we conclude that the connectome states are analogous to the classical gravity states. In the classical limit all Chern-Simons states that have diagrammatic presentation reduce to connectome states. Similarly, the holographic CFT states are those that have classical geometry interpretation in the classical limit. In the same spirit we can say that connectomes correspond to classical topologies. In both cases the entanglement entropy satisfies a minimal area law formula.

Unlike classical states quantum states are generally linear combinations of classical states. In particular, generic quantum states do not have a classical geometric, or topological interpretation. However there are instances of quantum states that do have classical limit different from that of connectome states. These are the states with nontrivial global three-dimensional topology as states~(\ref{holedstate}) and~(\ref{holedstate1}).  

In the classical limit states~(\ref{holedstate}) and~(\ref{holedstate1}) both reduce to the same state
\be
|00\rangle + \frac{1}{3}|11\rangle\,,
\ee
which is not a connectome state and has a non-minimal-area entropy~(\ref{topasympt}). Yet these states belong to the same Hilbert space as states~(\ref{chainedstate0}) and~(\ref{ringedstate}) and can be expressed as linear combinations of two connectomes. In other words, three-dimensional defects in topology introduce different classes of classical solutions.

\section{Wormholes}
\label{sec:wormholes}

In this section we will give further interpretation of the states with nontrivial three-dimensional topology. As we have seen, it is not very clear whether such states respect some form of the minimal area law (RT formula) for the entropy. Neither they reduce to the connectome states in the classical limit. We will explain how such states can be viewed as ``nonperturbative'', compared to the ``perturbative'' connectome states. 

The violation of the minimal area law is not very surprising -- similar problem occurs in the tensor network constructions~\cite{Pastawski:2015qua} -- but problematic, because we would like to use this law also for states with more than two $S^2$ boundaries. For instance, if we consider the case of three $S^2$, each with four punctures, and consider the following state,
\be
\begin{array}{c}
\begin{tikzpicture}[thick]
\fill[black] (0,0.0) circle (0.05cm);
\fill[black] (0,0.2) circle (0.05cm);
\fill[black] (0,0.4) circle (0.05cm);
\fill[black] (0,0.6) circle (0.05cm);
\draw (0,0) -- (1.8,0.0);
\fill[black] (1.8,0.0) circle (0.05cm);
\fill[black] (1.8,0.2) circle (0.05cm);
\fill[black] (1.8,0.4) circle (0.05cm);
\fill[black] (1.8,0.6) circle (0.05cm);
\draw (0,0.2) -- (1.8,0.2);
\fill[black] (0.6,1.2) circle (0.05cm);
\fill[black] (0.8,1.2) circle (0.05cm);
\fill[black] (1.,1.2) circle (0.05cm);
\fill[black] (1.2,1.2) circle (0.05cm);
\draw (0.6,1.2) -- (0.6,1.0) arc (0:-90:0.4) -- (0.,0.6);
\draw (0.8,1.2) -- (0.8,1.0) arc (0:-90:0.6) -- (0.,0.4);
\draw (1.0,1.2) -- (1.0,1.0) arc (180:270:0.6) -- (1.8,0.4);
\draw (1.2,1.2) -- (1.2,1.0) arc (180:270:0.4) -- (1.8,0.6);
\end{tikzpicture} 
\end{array} \ =\  |000\rangle + \frac{1}{\sqrt{d^2 - 1}} |111\rangle\,,
\ee
the reduced density matrix of any of the three spheres will be equivalent to the state~(\ref{holedstate}). In other words, it will contain a defect, so that the entanglement entropy will not be given by the minimal area law. We will show how the minimal area law is recovered if we ``patch'' the holes.

We first note that three-dimensional defects can be constructed via a \emph{surgery} operation on a trivial topology (e.g.~\cite{Witten:1988hf}). For example, $S^2\times S^1$ can be obtained by cutting a solid torus out of $S^3$, twisting the torus by exchanging its two fundamental cycles ($S$ modular transformation) and gluing the twisted torus with the remaining part of $S^3$, which in this case is also a solid torus. After such an operation, a noncontractible $S^1$ cycle appears in the resulting space in the place of the contractible cycle of the original torus. This is what diagrams in equations~(\ref{holedstate}) and~(\ref{holedstate1}) are intended to show.

Note that from the TQFT point of view, gluing to tori along a common boundary produces an overlap $\langle v_1|v_2\rangle$ of two vectors $|v_1\rangle$ and $|v_2\rangle$, each represented by a solid torus. If $S$ is an operator that exchanges the two fundamental cycles, then the result of the surgery is $\langle v_1|S|v_2\rangle$. 

$SU(2)$ Chern-Simons states with $T^2$ boundary have a special basis, labeled by integrable representations of $su(2)_k$ Kac-Moody algebra, which can be counted by Dynkin labels $R=0,\ldots, k$~\cite{Verlinde:1988sn,Witten:1988hf}. Such basis states are solid tori with a closed Wilson loop around the noncontractible cycle, colored with a given representation $R$. Matrix elements $S_{R_1R_2}=\langle R_1|S|R_2\rangle$ in this basis are computed by the $S^3$ invariants of  Hopf links with components colored by representations $R_1$ and $R_2$,\footnote{Here, again, we use a slightly different normalization from the standard one, to be consistent with the conventions chosen for the invariants. In the present normalization, the components $S_{0R}$ are the quantum dimensions of the representations, up to a sign, but matrix $S$ is not idempotent. }
\be
\begin{array}{c}
\begin{tikzpicture}[scale=0.7]
  \draw[line width=2.0,pink] (0,0) arc [x radius=1, y radius=0.3, start angle=-260, end angle=80];
  \draw[line width=2.0,cyan] (0.15,-0.45) arc [x radius=1, y radius=1, start angle=-20, end angle=320];
\end{tikzpicture} 
\end{array} = \ (-1)^{R_1+R_2}\frac{A^{2(R_1+1)(R_2+1)}-A^{-2(R_1+1)(R_2+1)}}{A^2-A^{-2}} \ = \ (-1)^{R_1+R_2}\frac{\sin\left(\frac{\pi(R_1+1)(R_2+1)}{k+2}\right)}{\sin\left(\frac{\pi}{k+2}\right)}\,.
\ee

Now, for state~(\ref{holedstate}), for example, we can write
\be
\label{blackhole}
\begin{array}{c}
    \begin{tikzpicture}[thick,xscale=cos(70)]
    \draw[double distance=1.4mm,opacity=0.5] (0:1) arc (-90:-270:0.3);
    \draw[rounded corners = 2] (-2,0.15) -- (-1.1,0.15) -- (-1.1,0.35) -- (3.1,0.35) -- (3.1,0.15) -- (4,0.15);
    \draw (-2,0.45) -- (4,0.45);
    \foreach \y in {0,0.1,...,0.2}
           \draw (0,0.15+\y) -- (2,0.15+\y);
    \draw[double distance=1.4mm,opacity=0.7] (0:1) arc (-90:90:0.3);
    \draw[rounded corners = 2] (2,0.15) -- (2.5,0.15) -- (2.5,-0.15) -- (-0.5,-0.15) -- (-0.5,0.15) -- (0,0.15);
    \draw[rounded corners = 2] (2,0.25) -- (2.8,0.25) -- (2.8,-0.25) -- (-0.8,-0.25) -- (-0.8,0.25) -- (0,0.25);
    \draw[rounded corners = 2] (-2,-0.15) -- (-1.1,-0.15) -- (-1.1,-0.35) -- (3.1,-0.35) -- (3.1,-0.15) -- (4,-0.15);
    \draw (-2,-0.45) -- (4,-0.45);
    \foreach \y in {0,0.3,...,1.2}
           \fill[black] (-2,-0.45+\y) ellipse (0.15cm and 0.05cm);
    \foreach \y in {0,0.3,...,1.2}
           \fill[black] (4,-0.45+\y) ellipse (0.15cm and 0.05cm);
  \end{tikzpicture}
\end{array} \ = \ \sum\limits_{R=0}^{k} S_{0R}^{-1} \begin{array}{c}
    \begin{tikzpicture}[thick,xscale=cos(70)]
    \draw[line width=0.7mm,opacity=0.2] (1,0.05) arc (-130:-220:0.35);
    \draw[rounded corners = 2] (-2,0.15) -- (-1.1,0.15) -- (-1.1,0.35) -- (3.1,0.35) -- (3.1,0.15) -- (4,0.15);
    \draw (-2,0.45) -- (4,0.45);
    \foreach \y in {0,0.1,...,0.2}
           \draw (0,0.15+\y) -- (2,0.15+\y);
    \draw[line width=0.7mm,gray] (1,0.05) arc (-130:150:0.35);
    \draw[line width=0.02mm,opacity=0.6] (1,0.05) arc (-130:150:0.35);
    \draw[line width=0.02mm,opacity=0.6] (1.07,0.05) arc (-130:150:0.35);
    \draw[line width=0.02mm,opacity=0.6] (0.93,0.05) arc (-130:150:0.35);
    \draw[rounded corners = 2] (2,0.15) -- (2.5,0.15) -- (2.5,-0.15) -- (-0.5,-0.15) -- (-0.5,0.15) -- (0,0.15);
    \draw[rounded corners = 2] (2,0.25) -- (2.8,0.25) -- (2.8,-0.25) -- (-0.8,-0.25) -- (-0.8,0.25) -- (0,0.25);
    \draw[rounded corners = 2] (-2,-0.15) -- (-1.1,-0.15) -- (-1.1,-0.35) -- (3.1,-0.35) -- (3.1,-0.15) -- (4,-0.15);
    \draw (-2,-0.45) -- (4,-0.45);
    \foreach \y in {0,0.3,...,1.2}
           \fill[black] (-2,-0.45+\y) ellipse (0.15cm and 0.05cm);
    \foreach \y in {0,0.3,...,1.2}
           \fill[black] (4,-0.45+\y) ellipse (0.15cm and 0.05cm);
    \draw[opacity=0.8] (0.4,0) node {\tiny $R$};
  \end{tikzpicture}
\end{array}\,.
\ee
In the right hand side, the nontrivial 3D topology was replaced by a sum of simpler 3D topologies, featuring 1D defects (loops) colored by representations $R$. Note that any representation $R=n>1$ can be obtained from $n$ copies of $R=1$ projecting out $R<n$ irreps from the tensor product. Diagrammatically this is achieved via the Jones-Wenzl projector (symmetrizer) \cite{Jones:1985dw,Wenzl:1985seq,Kauffman:2013bh}, which is a linear combination of elements of the Temperley-Lieb algebra of order $n$. To summarize the result of this operation without technical details we say that state~(\ref{holedstate}) is a linear combination of many diagrams in the trivial three-dimensional topology. 

In the classical limit $k\to\infty$, the number of diagram creating the defect is very large. If this limit is taken naively, one could expect that~(\ref{blackhole}) becomes
\be
\sum\limits_{R=0}^{k} S_{0R}^{-1} \begin{array}{c}
    \begin{tikzpicture}[thick,xscale=cos(70)]
    \draw[line width=0.7mm,opacity=0.2] (1,0.05) arc (-130:-220:0.35);
    \draw[rounded corners = 2] (-2,0.15) -- (-1.1,0.15) -- (-1.1,0.35) -- (3.1,0.35) -- (3.1,0.15) -- (4,0.15);
    \draw (-2,0.45) -- (4,0.45);
    \foreach \y in {0,0.1,...,0.2}
           \draw (0,0.15+\y) -- (2,0.15+\y);
    \draw[line width=0.7mm,gray] (1,0.05) arc (-130:150:0.35);
    \draw[line width=0.02mm,opacity=0.6] (1,0.05) arc (-130:150:0.35);
    \draw[line width=0.02mm,opacity=0.6] (1.07,0.05) arc (-130:150:0.35);
    \draw[line width=0.02mm,opacity=0.6] (0.93,0.05) arc (-130:150:0.35);
    \draw[rounded corners = 2] (2,0.15) -- (2.5,0.15) -- (2.5,-0.15) -- (-0.5,-0.15) -- (-0.5,0.15) -- (0,0.15);
    \draw[rounded corners = 2] (2,0.25) -- (2.8,0.25) -- (2.8,-0.25) -- (-0.8,-0.25) -- (-0.8,0.25) -- (0,0.25);
    \draw[rounded corners = 2] (-2,-0.15) -- (-1.1,-0.15) -- (-1.1,-0.35) -- (3.1,-0.35) -- (3.1,-0.15) -- (4,-0.15);
    \draw (-2,-0.45) -- (4,-0.45);
    \foreach \y in {0,0.3,...,1.2}
           \fill[black] (-2,-0.45+\y) ellipse (0.15cm and 0.05cm);
    \foreach \y in {0,0.3,...,1.2}
           \fill[black] (4,-0.45+\y) ellipse (0.15cm and 0.05cm);
    \draw[opacity=0.8] (0.4,0) node {\tiny $R$};
  \end{tikzpicture}
\end{array}
\ \to \ 
\left(\sum\limits_{R=0}^{\infty} S_{0R}^{-1} 
\begin{array}{c}
    \begin{tikzpicture}[thick,xscale=cos(70)]
    \draw[line width=0.7mm,opacity=0.6] (1,0.05) arc (-130:-220:0.35);
    \draw[line width=0.7mm,gray] (1,0.05) arc (-130:150:0.35);
    \draw[line width=0.02mm,opacity=0.6] (1,0.05) arc (-130:150:0.35);
    \draw[line width=0.02mm,opacity=0.6] (1.07,0.05) arc (-130:150:0.35);
    \draw[line width=0.02mm,opacity=0.6] (0.93,0.05) arc (-130:150:0.35);
    \draw[opacity=0.8] (0.4,0) node {\tiny $R$};
  \end{tikzpicture}
\end{array}\right)
\begin{array}{c}
    \begin{tikzpicture}[thick,xscale=cos(70)]
    \draw[rounded corners = 2] (-2,0.15) -- (-1.1,0.15) -- (-1.1,0.35) -- (3.1,0.35) -- (3.1,0.15) -- (4,0.15);
    \draw (-2,0.45) -- (4,0.45);
    \foreach \y in {0,0.1,...,0.2}
           \draw (0,0.15+\y) -- (2,0.15+\y);
    \draw[rounded corners = 2] (2,0.15) -- (2.5,0.15) -- (2.5,-0.15) -- (-0.5,-0.15) -- (-0.5,0.15) -- (0,0.15);
    \draw[rounded corners = 2] (2,0.25) -- (2.8,0.25) -- (2.8,-0.25) -- (-0.8,-0.25) -- (-0.8,0.25) -- (0,0.25);
    \draw[rounded corners = 2] (-2,-0.15) -- (-1.1,-0.15) -- (-1.1,-0.35) -- (3.1,-0.35) -- (3.1,-0.15) -- (4,-0.15);
    \draw (-2,-0.45) -- (4,-0.45);
    \foreach \y in {0,0.3,...,1.2}
           \fill[black] (-2,-0.45+\y) ellipse (0.15cm and 0.05cm);
    \foreach \y in {0,0.3,...,1.2}
           \fill[black] (4,-0.45+\y) ellipse (0.15cm and 0.05cm);
  \end{tikzpicture}
\end{array}
\ \sim \ |00\rangle + |11\rangle\,.
\ee
In other words, one could expect that in the classical limit the loops can be removed and one obtains a maximally entangled state. This is not quite the case, however. This operation does not simply commute with the summation in the formula. It is easy to check that for any finite $k$ the state is the one given by~(\ref{holedstate}), that is, it is not maximally entangled.

From the point of view of classical limit, Wilson loops are quantum fluctuations of the topology. Linear combination of an infinite number of diagrams, with many loops, creates a nonperturbative effect that persists in the classical limit in the form of a defect in the global topology. 

Let us discuss how the minimal area law works for these wormhole-like topologies. Let us consider a pair of two-spheres in $S^2\times S^3$ connected by $n+m$ lines, such $n$ and $m$ lines follow complementary segments of $S^1$, as the following diagram illustrates:
\be
\label{S2S1state}
\begin{array}{c}
    \begin{tikzpicture}[thick]
    \draw[line width=0.3,yscale=cos(50)] (1.1,0) arc (180:0:1.5);
    \draw[line width=0.3,yscale=cos(50)] (1,0) arc (180:0:1.6);
    \draw[line width=0.3,yscale=cos(50)] (0.9,0) arc (180:0:1.7);
    \draw[line width=0.3,yscale=cos(50)] (0.8,0) arc (180:0:1.8);
    \fill[gray] (1,0) circle (0.35);
     \fill[gray] (4.2,0) circle (0.35);
     \fill[black] (0.8,0) circle (0.05);
     \fill[black] (0.9,0) circle (0.05);
     \fill[black] (1.0,0) circle (0.05);
     \fill[black] (1.1,0) circle (0.05);
     \fill[black] (1.2,0) circle (0.05);
     \fill[black] (4.4,0) circle (0.05);
     \fill[black] (4.3,0) circle (0.05);
     \fill[black] (4.2,0) circle (0.05);
     \fill[black] (4.1,0) circle (0.05);
     \fill[black] (4.0,0) circle (0.05);
     \draw[line width=0.3,yscale=cos(50)] (0.8,0) arc (-180:0:1.8);
     \draw[line width=0.3,yscale=cos(50)] (0.9,0) arc (-180:0:1.7);
     \draw[line width=0.3,yscale=cos(50)] (1.,0) arc (-180:0:1.6);
     \draw[line width=0.3,yscale=cos(50)] (1.1,0) arc (-180:0:1.5);
     \draw[line width=0.3,yscale=cos(50)] (1.2,0) arc (-180:0:1.4);
    \draw[double distance=12mm,opacity=0.5,yscale=cos(50)] (0:1) arc (180:0:1.6);
    \draw[double distance=12mm,opacity=0.5,yscale=cos(50)] (0:1) arc (-180:0:1.6);
  \end{tikzpicture}
\end{array} 
\ee
Let us calculate the entanglement entropy corresponding to one of the spheres. There are two candidates for the minimal surface: one encircles the sphere and is crossed by $n+m$ lines; the other is a surface consistent of two disconnected $S^2$, one pierced with $m$ and the other with $n$ lines. The first surface (sphere) corresponds to a Hilbert space of dimension $C_{\frac{n+m}{2}}$, while the other to a product of two Hilbert spaces with total dimension $C_{\frac{n}{2}}\cdot C_{\frac{m}{2}}$. 

The dimension of the smaller Hilbert space defines the rank of the density matrix and consequently, the entanglement entropy. For finite $m$ and $n$ one has
\be
\label{Catalanineq}
C_{\frac{n+m}{2}}\  \geq  \ C_{\frac{n}{2}}\cdot C_{\frac{m}{2}}\,,
\ee
so the entanglement entropy, of one of the spheres in state~(\ref{S2S1state}) is $\log C_{\frac{n}{2}}\cdot C_{\frac{m}{2}}$. Indeed this is state is nothing but a projector of the Hilbert space of a sphere with $n+m$ punctures onto a subspace isomorphic to a product of two Hilbert spaces of spheres with $n$ punctures and $m$ punctures respectively. 

In the limit of very large $n$ and $m$, the leading order asymptotic of the Catalan number is $C_{\frac{n}{2}}\sim n\log 2$, so that~(\ref{Catalanineq}) gets saturated and the entropy calculation is consistent with counting the number of lines crossing the minimal surface, which is $m+n$ for both the connected and the disconnected surfaces. Essentially, for large number of degrees of freedom $n$ and $m$, the presence of the defect becomes unimportant, and this limit can be compared with the limit of low curvature of the holographic states.

Let us now review how the minimal area formula is recovered in the example of states  similar to~(\ref{holedstate}). In comparison to state~(\ref{S2S1state}) the analysis is complicated by the presence of additional Wilson loop winding the defect. For an arbitrary number $\ell$ of such Wilson loops the states would be
\be
\label{patchedstate}
C_{{\ell}/{2}}|00\rangle + \frac{C_{{\ell}/2}^{(1)}}{\sqrt{d^2-1}}|11\rangle\,,
\ee
where $C_{{\ell}/{2}}^{(j)}$ counts the multiplicity of the spin $j$ representation of $SU(2)$ in the tensor product of $\ell$ fundamental irreps,
\be
C_{{\ell}/{2}}^{(j)} \ = \ \frac{(2j+1)\ell!}{(\frac{\ell}{2}-j)!(\frac{\ell}{2}+j+1)!}\,.
\ee
Again, this counting is valid in the regime of large $k$. Note that when $\ell$ is also large, one has
\be
C_{{\ell}/{2}}^{(j)} \ \sim \ (2j+1)C_{{\ell}/{2}}\,, \qquad \ell\ \gg \ 1\,.
\ee
Consequently, state~(\ref{patchedstate}) becomes maximally entangled in the regime $k\gg \ell\gg 1$, 
\be
C_{{\ell}/{2}}|00\rangle + \frac{C_{{\ell}/2}^{(1)}}{\sqrt{d^2-1}}|11\rangle \ \to \  C_{{\ell}/{2}}\left(|00\rangle + |11\rangle \right) \,.
\ee

The presence of the large number of Wilson loops winding the defect restores the minimal area law in the sense discussed in this work, since in this case the minimal number of Wilson lines a surface separating two parts of the system can cross is four. 

This result can be generalized to any number of Wilson lines connecting the two parts. In particular, it solves the problem mentioned in the beginning of this section: as long as the number of Wilson lines is large the naive minimal area law formula counting the number of Wilson lines (connections between the parties) is valid. In particular, this implies that connectome states for large $k$ and large number of Wilson lines satisfy the inequalities for the entanglement entropy, satisfied by the holographic states, as discussed in~\cite{Melnikov:2023nzn}.

\section{Replica wormholes and ordinary replicas}
\label{sec:replicawormholes}

In this section we use the analogy between the connectome and holographic states to revisit the problem of the replica wormholes appearing in the discussion of the black hole information paradox. We will try to construct analogs of replica wormholes in terms of the Chern-Simons theory, or a more general TQFT, and observe that such objects are not unique to gravity theories.

In the gravitational context replica wormholes appear after one chooses the following prescription for the calculation of the path integral: fix boundary conditions and sum over all geometries and topologies~\cite{Penington:2019kki,Almheiri:2019qdq}. When applied to a replicated manifold in the replica method this prescription might be understood as a requirement of including all possible topologies in consideration, even those that are not factorizable. One way the problem may be stated heuristically is $\rho^n\neq(\rho)^n$, where $\rho$ is the density matrix, or path integral being replicated.

Since the problem appears in the part of the calculation that deals with the sum over topologies, one might expect to see an analog of this problem in a Chern-Simons calculation. We will not, however, observe a necessity to sum over all possible topologies in Chern-Simons path integral. Since all the calculations can be performed exactly, the correct answers are obtained without including all replica wormholes, but rather some specific ones. We will conclude that, somewhat tautologically, topologies which appear as alternative saddles in the replica path integral must transform into each other through an appropriate time evolution.

Hawking's original calculation~\cite{Hawking:1976ra} of the entropy of the black hole radiation predicts a monotonous growth of this entropy until the moment black hole evaporates completely, producing a mixed final state from a pure original one, contradicting unitarity. Page~\cite{Page:1993wv} assumed instead that the evolution is unitary and the state of the full system remains pure through the process, which implies that the entropy of the radiation must vanish at the end of the evaporation. In other words, some late time effect should modify Hawking's prediction, leading to zero entropy in the final state. Recently it was found~\cite{Penington:2019npb,Almheiri:2019psf,Almheiri:2019hni} that the generalized entropy involving quantum extremal surfaces~\cite{Engelhardt:2014gca} reproduces the behavior of the entropy predicted by Page. It was then argued~\cite{Penington:2019kki,Almheiri:2019qdq} that the decaying behavior of the entropy at late times is due to the dominance of nontrivial topologies in the path integral -- replica wormholes.

Hawking's calculation deals with quantum fields in a fixed gravitational background. Its result can be compared with the analysis in section~\ref{sec:minarrea}, where a power of the density matrix, which can be thought as of density matrix of the radiation is computed by a simple concatenation, cf.~(\ref{denmatrix}). Let us schematically denote the state of the entangled Hawking pairs as
\be
|\Psi\rangle \ = \ \begin{array}{c}
     \begin{tikzpicture}
         \draw[line width=2,orange] (0,0) -- (1.,0);
         \draw[line width=2,black] (2,0) -- (1,0);
         \fill[orange] (0,0) circle (0.1);
         \fill[black] (2,0) circle (0.1);
     \end{tikzpicture} 
\end{array}, 
\ee
where the black dot represents quanta in the interior. Then, the traces of the reduced density matrix of the radiation and its powers are given by
\be
\label{Hawkingcalc}
\tr\rho \ \sim \ \begin{array}{c}
     \begin{tikzpicture}
         \draw[line width=2,orange] (1,0) -- (0.5,0) arc (270:90:0.25) -- (1,0.5);
         \draw[line width=2,black] (1,0) -- (1.5,0) arc (-90:90:0.25) -- (1,0.5);
     \end{tikzpicture} 
\end{array}\,, \qquad 
\tr\rho^2 \ \sim \ \begin{array}{c}
     \begin{tikzpicture}
         \draw[line width=2,orange] (1,0) -- (0.5,0) arc (270:90:0.25) -- (2,0.5) arc (90:-90:0.25) -- (1,0);
         \foreach \y in {0,0.5,...,0.5}
           \draw[line width=2,black] (1,0+\y) -- (1.5,0+\y);
     \end{tikzpicture} 
\end{array}
\,, \qquad 
\tr\rho^3 \ \sim \ \begin{array}{c}
     \begin{tikzpicture}
         \draw[line width=2,orange] (2,0) -- (0.5,0) arc (270:90:0.25) -- (2,0.5);
         \foreach \y in {0,0.5,...,0.5}
           \draw[line width=2,black] (1,0+\y) -- (1.5,0+\y);
           \draw[line width=2,black] (2,0) arc (-90:90:0.25);
     \end{tikzpicture} 
\end{array} \ldots
\ee
In accordance with the discussion in the earlier parts of this work, such replicas would count the quanta entangled across the horizon in the classical limit. Consequently, the entropy will increase monotonously with the number of emitted quanta.

Replica wormholes are topologies different from those appearing in~(\ref{Hawkingcalc}). They are said to correspond to topologies connecting different replicas. Naively, one can imagine that a generic replica wormhole topology is something like schematically illustrated by the following diagrams (here we consider examples of $n=2$ and $n=4$):
\be
\label{replicaworm}
\begin{array}{c}
     \begin{tikzpicture}
         \draw[line width=2,orange] (3.0,0) -- (3.5,0) arc (-90:90:0.5) -- (3.0,1);
         \draw[line width=2,orange] (1,0) -- (0.5,0) arc (270:90:0.5) -- (1,1);
           \draw[line width=2,black,opacity=0.5] (1,0) -- (3,0) arc (270:90:0.5) -- (1,1) arc (90:-90:0.5);
     \end{tikzpicture} 
\end{array}\,, \qquad
\begin{array}{c}
     \begin{tikzpicture}
         \draw[line width=2,orange] (3.5,0) -- (0.5,0) arc (270:90:0.5) -- (3.5,1) arc (90:-90:0.5);
         \foreach \y in {0,1,...,1}
           \draw[line width=2,white] (1,0+\y) -- (1.75,0+\y);
        \foreach \y in {0,1,...,1}
           \draw[line width=2,white] (2.25,0+\y) -- (3.0,0+\y);
           \foreach \y in {0,1,...,1}
           \draw[line width=2,black,opacity=0.3] (1,0+\y) -- (1.75,0+\y);
        \foreach \y in {0,1,...,1}
           \draw[line width=2,black,opacity=0.3] (2.25,0+\y) -- (3.0,0+\y);
        \draw[line width=2,black,rounded corners=2,opacity=0.3] (1.,0) -- (1.25,0) -- (1.25,1) -- (1,1);
        \draw[line width=2,black,rounded corners=2,opacity=0.3] (2.25,0) -- (2.5,0) -- (2.5,1) -- (2.25,1);
        \draw[line width=2,black,rounded corners=2,opacity=0.3] (1.75,0) -- (1.5,0) -- (1.5,1) -- (1.75,1);
        \draw[line width=2,black,rounded corners=2,opacity=0.3] (3.0,0) -- (2.75,0) -- (2.75,1) -- (3.,1);
        \draw[line width=2,rounded corners=2,black,opacity=0.3] (2.25,0) -- (2.25,0.15) -- (1.75,0.15) -- (1.75,0.);
        \draw[line width=2,rounded corners=2,black,opacity=0.3]  (3.0,0) -- (3.0,0.25) -- (1.,0.25) -- (1.,0.);
        \draw[line width=2,rounded corners=2,black,opacity=0.3] (2.25,1) -- (2.25,0.85) -- (1.75,0.85) -- (1.75,1);
        \draw[line width=2,rounded corners=2,black,opacity=0.3]  (3.0,1) -- (3.0,0.75) -- (1.,0.75) -- (1.,1);
        \draw[line width=2,rounded corners=2,black,opacity=0.3]  (3.0,0) -- (3.0,0.65) -- (1.75,0.65) -- (1.75,1);
        \draw[line width=2,rounded corners=2,black,opacity=0.3]  (1.0,0) -- (1.0,0.35) -- (2.25,0.35) -- (2.25,1);
        \draw[line width=2,rounded corners=2,black,opacity=0.3]  (2.25,0) -- (2.35,0.) -- (2.35,0.45) -- (1,0.45) -- (1,1);
        \draw[line width=2,rounded corners=2,black,opacity=0.3]  (3,1) -- (2.9,1) -- (2.9,0.55) -- (1.75,0.55) -- (1.75,0);
     \end{tikzpicture} 
\end{array}\,.
\ee
In a generic case, like shown above, the beginning (end) of each replica is connected by some number of links with the ends (beginnings) of all the remaining replicas. Now we are going to argue that generic replica wormholes should not appear in the calculation.

Chern-Simons theory per se, does not have a dynamical Hamiltonian that could evolve one entangled state into another. So the situation is more similar to Hawking's fixed background scenario. On the other hand, recently proposed models, which consider replica wormholes, do have a dynamical evolution, in which the black hole degrees of freedom dissipate energy into radiation~\cite{Almheiri:2019qdq}. After all, this dissipation is a transfer of quanta from the interior to the exterior. In terms of the discussion in this work this process can be illustrated as follows:
\begin{eqnarray}
\begin{array}{c}
\begin{tikzpicture}[thick]
\newcommand{\x}{1.5}
\newcommand{\z}{0}
\foreach \y in {0.0,0.3,...,1.0}
            \draw (0+\z,\y) -- (0.4+\z,\y) arc (-90:90:0.1cm) -- (0+\z,\y+0.2);
\foreach \y in {0,0.3,...,1}
           \fill[black] (0+\z,\y) circle (0.04cm);
\foreach \y in {0.2,0.5,...,1.4}
           \fill[black] (0+\z,\y) circle (0.04cm);
\foreach \y in {0,0.3,...,1}
           \draw[gray] (\x+\z,\y) circle (0.04cm);
\foreach \y in {0.2,0.5,...,1.4}
           \draw[gray] (\x+\z,\y) circle (0.04cm);
\end{tikzpicture} 
\end{array}  \qquad  & \qquad  \begin{array}{c}
\begin{tikzpicture}[thick]
\newcommand{\x}{1.5}
\newcommand{\z}{0}
\foreach \y in {0.3,0.6,...,0.7}
            \draw (0+\z,\y) -- (0.4+\z,\y) arc (-90:90:0.1cm) -- (0+\z,\y+0.2);
\foreach \y in {0,0.2,...,0.2}
            \draw (0+\z,\y) -- (\x+\z,\y);
\foreach \y in {0,0.3,...,0.7}
           \fill[black] (0+\z,\y) circle (0.04cm);
\foreach \y in {0.9,1.1,...,1.1}
           \draw[gray] (0+\z,\y) circle (0.04cm);
\foreach \y in {0.2,0.5,...,1.0}
           \fill[black] (0+\z,\y) circle (0.04cm);
\foreach \y in {0.3,0.6,...,1}
           \draw[gray] (\x+\z,\y) circle (0.04cm);
\foreach \y in {0.5,0.8,...,1.4}
           \draw[gray] (\x+\z,\y) circle (0.04cm);
\foreach \y in {0.0,0.2,...,0.2}
           \fill[black] (\x+\z,\y) circle (0.04cm);
\end{tikzpicture} 
\end{array} \qquad   & \qquad \begin{array}{c}
\begin{tikzpicture}[thick]
\newcommand{\x}{-1.5}
\newcommand{\z}{1.5}
\foreach \y in {0.3,0.6,...,0.7}
            \draw (0+\z,\y) -- (-0.4+\z,\y) arc (270:90:0.1cm) -- (0+\z,\y+0.2);
\foreach \y in {0,0.2,...,0.2}
            \draw (0+\z,\y) -- (\x+\z,\y);
\foreach \y in {0,0.3,...,0.7}
           \fill[black] (0+\z,\y) circle (0.04cm);
\foreach \y in {0.9,1.1,...,1.1}
           \draw[gray] (0+\z,\y) circle (0.04cm);
\foreach \y in {0.2,0.5,...,1.0}
           \fill[black] (0+\z,\y) circle (0.04cm);
\foreach \y in {0.3,0.6,...,1}
           \draw[gray] (\x+\z,\y) circle (0.04cm);
\foreach \y in {0.5,0.8,...,1.4}
           \draw[gray] (\x+\z,\y) circle (0.04cm);
\foreach \y in {0.0,0.2,...,0.2}
           \fill[black] (\x+\z,\y) circle (0.04cm);
\end{tikzpicture} 
\end{array}    \label{bhevolution}\\
\text{initial state}\qquad  & \qquad\text{early time} \qquad &\qquad \quad \text{late time} \nn
\end{eqnarray}
This is equivalent to moving the partition interior-exterior, what happens in the generalized entropy prescription, which moves the quantum extremal surfaces. The prescription itself does not provide the Hamiltonian, but certain models, at least in low dimensions, determine the position of the extremal surface dynamically~\cite{Almheiri:2019qdq}.

It is concluded that the nontrivial extremal surface corresponds to a classical geometry whose topology is a replica wormhole. The diagrams in formula~(\ref{bhevolution}) show, however, that what looks like a wormhole is in fact an ordinary replicated time evolved density matrix. Indeed, if one compares the reduced density matrices of the radiation (right hand side degrees of freedom in~(\ref{bhevolution})) of the early and late time states one will find something of the following kind:

\begin{eqnarray}
\begin{array}{c}
\begin{tikzpicture}[thick]
\newcommand{\x}{-1.5}
\newcommand{\z}{0}
\foreach \y in {0.3,0.6,...,0.7}
            \draw (0+\z,\y) -- (-0.4+\z,\y) arc (270:90:0.1cm) -- (0+\z,\y+0.2);
\foreach \y in {0,0.2,...,0.2}
            \draw (0+\z,\y) -- (\x+\z,\y);
\foreach \y in {0,0.3,...,0.7}
           \fill[black] (0+\z,\y) circle (0.04cm);
\foreach \y in {0.9,1.1,...,1.1}
           \draw[gray] (0+\z,\y) circle (0.04cm);
\foreach \y in {0.2,0.5,...,1.0}
           \fill[black] (0+\z,\y) circle (0.04cm);
\foreach \y in {0.3,0.6,...,1}
           \draw[gray] (\x+\z,\y) circle (0.04cm);
\foreach \y in {0.5,0.8,...,1.4}
           \draw[gray] (\x+\z,\y) circle (0.04cm);
\foreach \y in {0.0,0.2,...,0.2}
           \fill[black] (\x+\z,\y) circle (0.04cm);
\renewcommand{\x}{1.5}
\renewcommand{\z}{0}
\foreach \y in {0.3,0.6,...,0.7}
            \draw (0+\z,\y) -- (0.4+\z,\y) arc (-90:90:0.1cm) -- (0+\z,\y+0.2);
\foreach \y in {0,0.2,...,0.2}
            \draw (0+\z,\y) -- (\x+\z,\y);
\foreach \y in {0,0.3,...,0.7}
           \fill[black] (0+\z,\y) circle (0.04cm);
\foreach \y in {0.9,1.1,...,1.1}
           \draw[gray] (0+\z,\y) circle (0.04cm);
\foreach \y in {0.2,0.5,...,1.0}
           \fill[black] (0+\z,\y) circle (0.04cm);
\foreach \y in {0.3,0.6,...,1}
           \draw[gray] (\x+\z,\y) circle (0.04cm);
\foreach \y in {0.5,0.8,...,1.4}
           \draw[gray] (\x+\z,\y) circle (0.04cm);
\foreach \y in {0.0,0.2,...,0.2}
           \fill[black] (\x+\z,\y) circle (0.04cm);
\end{tikzpicture} 
\end{array} \qquad   & \text{and}& \qquad \begin{array}{c}
\begin{tikzpicture}[thick]
\newcommand{\x}{-1.5}
\newcommand{\z}{1.5}
\foreach \y in {0.3,0.6,...,0.7}
            \draw (0+\z,\y) -- (-0.4+\z,\y) arc (270:90:0.1cm) -- (0+\z,\y+0.2);
\foreach \y in {0,0.2,...,0.2}
            \draw (0+\z,\y) -- (\x+\z,\y);
\foreach \y in {0,0.3,...,0.7}
           \fill[black] (0+\z,\y) circle (0.04cm);
\foreach \y in {0.9,1.1,...,1.1}
           \draw[gray] (0+\z,\y) circle (0.04cm);
\foreach \y in {0.2,0.5,...,1.0}
           \fill[black] (0+\z,\y) circle (0.04cm);
\foreach \y in {0.3,0.6,...,1}
           \draw[gray] (\x+\z,\y) circle (0.04cm);
\foreach \y in {0.5,0.8,...,1.4}
           \draw[gray] (\x+\z,\y) circle (0.04cm);
\foreach \y in {0.0,0.2,...,0.2}
           \fill[black] (\x+\z,\y) circle (0.04cm);
\renewcommand{\x}{1.5}
\renewcommand{\z}{-1.5}
\foreach \y in {0.3,0.6,...,0.7}
            \draw (0+\z,\y) -- (0.4+\z,\y) arc (-90:90:0.1cm) -- (0+\z,\y+0.2);
\foreach \y in {0,0.2,...,0.2}
            \draw (0+\z,\y) -- (\x+\z,\y);
\foreach \y in {0,0.3,...,0.7}
           \fill[black] (0+\z,\y) circle (0.04cm);
\foreach \y in {0.9,1.1,...,1.1}
           \draw[gray] (0+\z,\y) circle (0.04cm);
\foreach \y in {0.2,0.5,...,1.0}
           \fill[black] (0+\z,\y) circle (0.04cm);
\foreach \y in {0.3,0.6,...,1}
           \draw[gray] (\x+\z,\y) circle (0.04cm);
\foreach \y in {0.5,0.8,...,1.4}
           \draw[gray] (\x+\z,\y) circle (0.04cm);
\foreach \y in {0.0,0.2,...,0.2}
           \fill[black] (\x+\z,\y) circle (0.04cm);
\end{tikzpicture} 
\end{array}\,.   
\end{eqnarray}

One should focus on the structure of the open ends of these density matrices. At early times, what one would see, is a classical topology corresponding to a (maximally) entangled configuration with entropy growing with time, as loops in the middle open and reconnect with the boundaries. At late times one would see more open ends, some of which, when concatenated, appearing like replica wormholes connecting the replicas in a cyclic fashion preserving the replica symmetry. Specifically, for $n=2$, one would find the same picture as illustrated by the first diagram in formula~(\ref{replicaworm}).

In the picture just described the transition between early and late times occurs continuously via the same dynamics moving endpoints of the lines from one side to the other. The precise dynamics may be complicated. In the context of topological theories one can imagine the dynamics introducing nonlocal braiding of the lines, which also affects the entropy~\cite{Melnikov:2022vij}. So in general, the quantities computing the entropy will be calculated by sums over different topologies. As an example, lets us think of state~(\ref{chainedstate0}). Any power of the reduced density matrix of this state can be expanded as
\be
\left(\begin{array}{c}
\begin{tikzpicture}[thick]
\newcommand{\x}{0.6}
\draw (0,0) -- (1.8,0);
\draw (0,0.3) -- (1.8,0.3);
\draw (0,0.6) -- (0.5,0.6) arc (-90:30:0.15cm);
\draw (0,0.9) -- (0.5,0.9) arc (90:60:0.15cm);
\draw (0.9,0.6) -- (0.7,0.6) arc (-90:-120:0.15cm);
\draw (0.9,0.6) -- (1.1,0.6) arc (-90:-60:0.15cm);
\draw (0.9,0.9) -- (0.7,0.9) arc (90:210:0.15cm);
\draw (0.9,0.9) -- (1.1,0.9) arc (90:-30:0.15cm);
\draw (1.2+\x,0.9) -- (0.7+\x,0.9) arc (90:120:0.15cm);
\draw (1.2+\x,0.6) -- (0.7+\x,0.6) arc (-90:-210:0.15cm);
\fill[black] (1.2+\x,0.0) circle (0.05cm);
\fill[black] (1.2+\x,0.3) circle (0.05cm);
\fill[black] (1.2+\x,0.6) circle (0.05cm);
\fill[black] (1.2+\x,0.9) circle (0.05cm);
\fill[black] (0,0.0) circle (0.05cm);
\fill[black] (0,0.3) circle (0.05cm);
\fill[black] (0,0.6) circle (0.05cm);
\fill[black] (0,0.9) circle (0.05cm);
\end{tikzpicture} 
\end{array}\right)^n \ = \ \alpha
\begin{array}{c}
\begin{tikzpicture}[thick]
\fill[black] (0,0.0) circle (0.05cm);
\fill[black] (0,0.3) circle (0.05cm);
\fill[black] (0,0.6) circle (0.05cm);
\fill[black] (0,0.9) circle (0.05cm);
\draw (0,0) -- (0.4,0) -- (1.2,0.0);
\draw (0,0.6) -- (0.4,0.6) arc (-90:90:0.15cm) -- (0,0.9);
\fill[black] (1.2,0.0) circle (0.05cm);
\fill[black] (1.2,0.3) circle (0.05cm);
\fill[black] (1.2,0.6) circle (0.05cm);
\fill[black] (1.2,0.9) circle (0.05cm);
\draw (1.2,0.3) -- (0,0.3);
\draw (1.2,0.6) -- (0.8,0.6) arc (-90:-270:0.15cm) -- (1.2,0.9);
\end{tikzpicture} 
\end{array}
 +  \beta  
\begin{array}{c}
\begin{tikzpicture}[thick]
\draw (0,0) -- (1.2,0);
\draw (0,0.6) -- (1.2,0.6);
\draw (1.2,0.3) -- (0,0.3);
\draw (1.2,0.9) -- (0,0.9);
\fill[black] (0,0.0) circle (0.05cm);
\fill[black] (0,0.3) circle (0.05cm);
\fill[black] (0,0.6) circle (0.05cm);
\fill[black] (0,0.9) circle (0.05cm);
\fill[black] (1.2,0.0) circle (0.05cm);
\fill[black] (1.2,0.3) circle (0.05cm);
\fill[black] (1.2,0.6) circle (0.05cm);
\fill[black] (1.2,0.9) circle (0.05cm);
\end{tikzpicture} 
\end{array}\,,
\ee
with some easily computable coefficients $\alpha$ and $\beta$. The result is a sum of connected and disconnected topologies. Moreover, the fact that the disconnected topology is $n$-factorizable is not very obvious without some additional information about the nature of these quantum states. A purifying evolution must be some nonlocal operation, which guaranties that the weight of the disconnected topology dominates at late times.

\section{Conclusions}
\label{sec:conclusions}

In this work we revisited construction and  properties of quantum states in Chern-Simons theory with $SU(2)$ gauge group. We used axiomatic TQFT approach, to represent quantum states by topological spaces and to develop intuition about their properties from the topological properties of spaces, such as connectivity.

As a paradigmatic example of quantum states we considered a class characterized by 3-manifolds limited by a pair of $S^2$ boundaries. The space connectivity was controlled by the wiring of Wilson lines connecting the punctures on the boundary spheres, closed Wilson loops in the bulk and global three-dimensional defects in the manifold topology. 

We studied different examples of quantum states in this family, illustrating how quantum entanglement depends on the way the space is connected. A general lesson learned about all the examples is that a lack of connectivity of space results in a lack of quantum entanglement between the subsystems, where by lack of connectivity we mean not only truly disconnected topologies, but also topologies with a complex connectivity. In other words, more holes -- less entanglement.

A central role in our discussion was played by the connectome states, which were introduced as states, or equivalence classes of states, with the simplest possible topology in a given finite-dimensional Hilbert space. In a bipartite system, such as the one characterized by two disconnected spacetime boundaries the connectome states correspond to planar graphs with two vertices, match the SLOCC classes of bipartite entanglement and represent states with the maximal entanglement in each class. They also provide a basis for all other Chern-Simons states. 

Our primary interest in the connectomes was due to their similarity with the holographic states. Their first relation is through the discrete analog of the RT formula for the entanglement entropy satisfied by the connectome states. In the present context the RT formula states that the entanglement entropy of a subsystem (one of the boundaries) is given by the logarithm of the dimension of the Hilbert space associated with a closed bulk surface encircling the subsystem in such a way that it is crossed by the minimum number of Wilson lines. In the special case, when the number of the Wilson lines is very large, the entanglement entropy is approximately equal to the number of lines times $\log 2$.

The second relation with holography is the fact that in the classical limit, all quantum state of Chern-Simons with global $S^3$ topology reduce to the connectome states. So connectomes are the classical topologies of Chern-Simons theory, as holographic states are those dual to classical geometries. In fact classical limit can be used as a definition of the connectome states. 

A more fundamental relation of the connectomes and the holographic states is the conjecture discussed in~\cite{Melnikov:2023nzn} that the connectomes satisfy the same inequalities for the entanglement entropy as the holographic states. We do not discuss the details of the inequalities here, but we clarify that the discussion of the inequalities is valid in the double limit, with large Chern-Simons coupling $k$ and large number of Wilson lines.

We also observed that states with defects in the global topology, or equivalently, states with global topology different from $S^3$, do not reduce to connectome states in the classical limit. We advocated that such states can be viewed as nonperturbative topologies compared to perturbative connectomes. Using surgery we showed that states with defects are ordinary connectome states with a very large number of quantum fluctuations that ``condense'' producing defects in the classical limit. 

We discussed how the RT formula works in the presence of defects, since the definition of the formula given above does not always produce the expected result. We showed that the RT formula works if the defects also have a large number of Wilson lines wound around them. This is also reminiscent of the situation in holography, where for classical formulas to work the space needs to have small curvature (or the dual theory to have very large number of degrees of freedom).  

We expect that the ideas and results of this work are useful in the discussion of the emergence of spacetime from entanglement. Here both entanglement and connectivity are supported by the Wilson lines linking different parts of the system. Wilson loops and nontrivial tangling of the Wilson lines appear as quantum fluctuations of the spacetime, and when such fluctuations are many, they can collapse and form an object modifying spacetime topology, like a black holes or a wormhole.

We also hope that the TQFT approach or intuition can help in the study of quantum gravity effects. Here we attempted to understand whether such objects as replica wormholes exist beyond the gravitational theories. Our conclusion was that replica wormholes are quite general and we find analogs of them in Chern-Simons. Our analysis suggests that contrary to some statements that path integral must include contribution of all possible topologies, only special topologies contribute. In particular, replica symmetry must be present in the full path integral. Which replica wormholes contribute at different stage of the evolution depends on the specific Hamiltonian. In the context of the black hole information paradox this means that replica wormholes by themselves cannot be a solution, without the knowledge of the Hamiltonian.  

\bigskip

\paragraph{Acknowledgments} The author would like to thank the ICTP-SAIFR for support and hospitality during the workshop Holography@25 and the Isaac Newton Institute for Mathematical Sciences, Cambridge, for support and hospitality during the programme "Black holes: bridges between number theory and holographic quantum information", where work on this paper was undertaken. This work was supported in part by Simons Foundation award number 1023171-RC, grants of the Brazilian National Council for Scientific and Technological Development (CNPq) number 308580/2022-2 and 404274/2023-4, grant of the Serrapilheira Institute number Serra R-2012-38185 and the EPSRC grant number EP/R014604/1.

\bibliographystyle{JHEP}
\bibliography{refs}

\end{document}